\documentclass[sigconf, 10pt, table]{acmart}

\AtBeginDocument{%
  }

\usepackage{subcaption}
\usepackage{bm}
\usepackage{ragged2e}
\usepackage{enumitem}
\usepackage{soul}
\usepackage{array}
\renewcommand\footnotetextcopyrightpermission[1]{} 

\setcopyright{none}

\usepackage{subcaption}
\usepackage{booktabs}
\usepackage{balance}

\usepackage{natbib}

\acmConference[]{}
\acmYear{}
\copyrightyear{}

\acmPrice{}

\usepackage{xspace}
\newcommand{\eg}{\textit{e.g.},\xspace}
\newcommand{\ie}{\textit{i.e.},\xspace}

\newcommand{\data}{\texttt{NetMob23}\xspace}

\newcommand{\rotate}[1]{\rotatebox[origin=lb]{90}{#1}}
\definecolor{Gray}{gray}{0.93}
\hypersetup{
    colorlinks=true,
    linkcolor=blue,
    filecolor=magenta,      
    urlcolor=cyan,
    pdftitle={Overleaf Example},
    pdfpagemode=FullScreen,
}

\begin{document}

\title[The NetMob23 Dataset]{\huge The NetMob23 Dataset: A High-resolution Multi-region Service-level Mobile Data Traffic Cartography}


\author[O.E. Mart\'inez-Durive, S. Mishra, C. Ziemlicki, S. Rubrichi, Z. Smoreda,  M. Fiore]{
Orlando E. Mart\'inez-Durive$^{\star\dag\ast}$, 
Sachit Mishra$^{\star\dag\ast}$,
Cezary Ziemlicki$^\ddagger$,\\
Stefania Rubrichi$^\ddagger$,
Zbigniew Smoreda$^\ddagger$,
and Marco Fiore$^{\star}$}
\affiliation{
$^\star$IMDEA Networks Institute \country{Spain};
$^\dag$Universidad Carlos III de Madrid \country{Spain};
$^\ddagger$Orange Innovation \country{France}.}
\email{ {orlando.martinez, sachit.mishra, marco.fiore}@imdea.org} \email{ {cezary.ziemlicki, stefania.rubrichi, zbigniew.smoreda}@orange.com }
\thanks{$\ast$ Equal contributors.}

\begin{abstract}
Digital sources have been enabling unprecedented data-driven and large-scale investigations across a wide range of domains, including demography, sociology, geography, urbanism, criminology, and engineering. A major barrier to innovation is represented by the limited availability of dependable digital datasets, especially in the context of data gathered by mobile network operators or service providers, due to concerns about user privacy and industrial competition. The resulting lack of reference datasets curbs the production of new research methods and results, and prevents verifiability and reproducibility of research outcomes.
The \data dataset offers a rare opportunity to the multidisciplinary research community to access rich data about the spatio-temporal consumption of mobile applications in a developed country. The generation process of the dataset sets a new quality standard, leading to information about the demands generated by $68$ popular mobile services, geo-referenced at a high resolution of $100\times100$ m\textsuperscript{2} over $20$ metropolitan areas in France, and monitored during $77$ consecutive days in $2019$.  
\end{abstract}

\maketitle

\vspace*{-8pt}
\section{Introduction}

The surge in the usage of mobile devices and Internet services is generating an enormous amount of data, which has a high and largely untapped potential to support the discovery of new knowledge about human behaviors. The data that can be collected in production mobile networks is already proving an invaluable proxy to analyze the habits of large populations at large scales of cities or countries, complementing and in some cases replacing traditional sources such as surveys or censuses that are expensive and time-consuming to run. Examples of the substantial utility of mobile network data for research abound, and span a plethora of domains: the data can unlock analyses of mobility patterns~\cite{gonzalez2008understanding,song2010limits,csaji2013exploring,kung2014exploring,louail2015uncovering} and social interactions~\cite{miritello2013limited}, explorations of transportation systems~\cite{seppecher21zonal} estimates of static and dynamic population density~\cite{deville2014dynamic,lenormand2015comparing,khodabandelou2018estimation,batista2020uncovering}, predictions of poverty~\cite{steele17poverty,pnas_17_combining_data_sources}, socioeconomic inequality~\cite{moro21atlas,ucar21news} or digital divides~\cite{mishra2022second},
and mappings of land usage~\cite{toole12landuse,lenormand15landuse,grauwin15cities,furno17spatiotemporal} or urban transformation~\cite{www_16_italian_cities_daniel_quercia} or pollution~\cite{chen21pollution};
in addition, mobile network data have proven instrumental to assess the impacts of natural disasters~\cite{yabe22disasters} or infectious disease transmission~\cite{www_19_carlos_serraute, oliver20covid,zanella2022impact}, and the effectiveness of the associated containment policies~\cite{heroy21covidpolicy,pullano2020evaluating}.

These data can also enable studies aimed at understanding how the mobile network infrastructure is used, improving its management and extend its functionalities: for instance, they can be leveraged for the localization and tracking of devices~\cite{www_16_linking_users,IMWUT_19_cellsense}, the characterization of network loads~\cite{paul2011infocom,www_20_cellrep} and application usages~\cite{marquez17conext}, the prediction of traffic fluctuations~\cite{zhang18mobihoc}, the planning~\cite{su22planning} and dynamic reconfiguration~\cite{sharma22vran} of Radio Access Networks (RAN) infrastructures, or the data-driven management of network resources in sliced environments~\cite{zhang20mobicom}.
The list above only covers a small sample of the literature and does not pretend to be comprehensive by any means; surveys are available that provide a more thorough review of the relevant body of works~\cite{blondel2015survey,naboulsi15survey}.

\begin{figure*}[tb]
\subfloat{\includegraphics[width=0.75\textwidth]{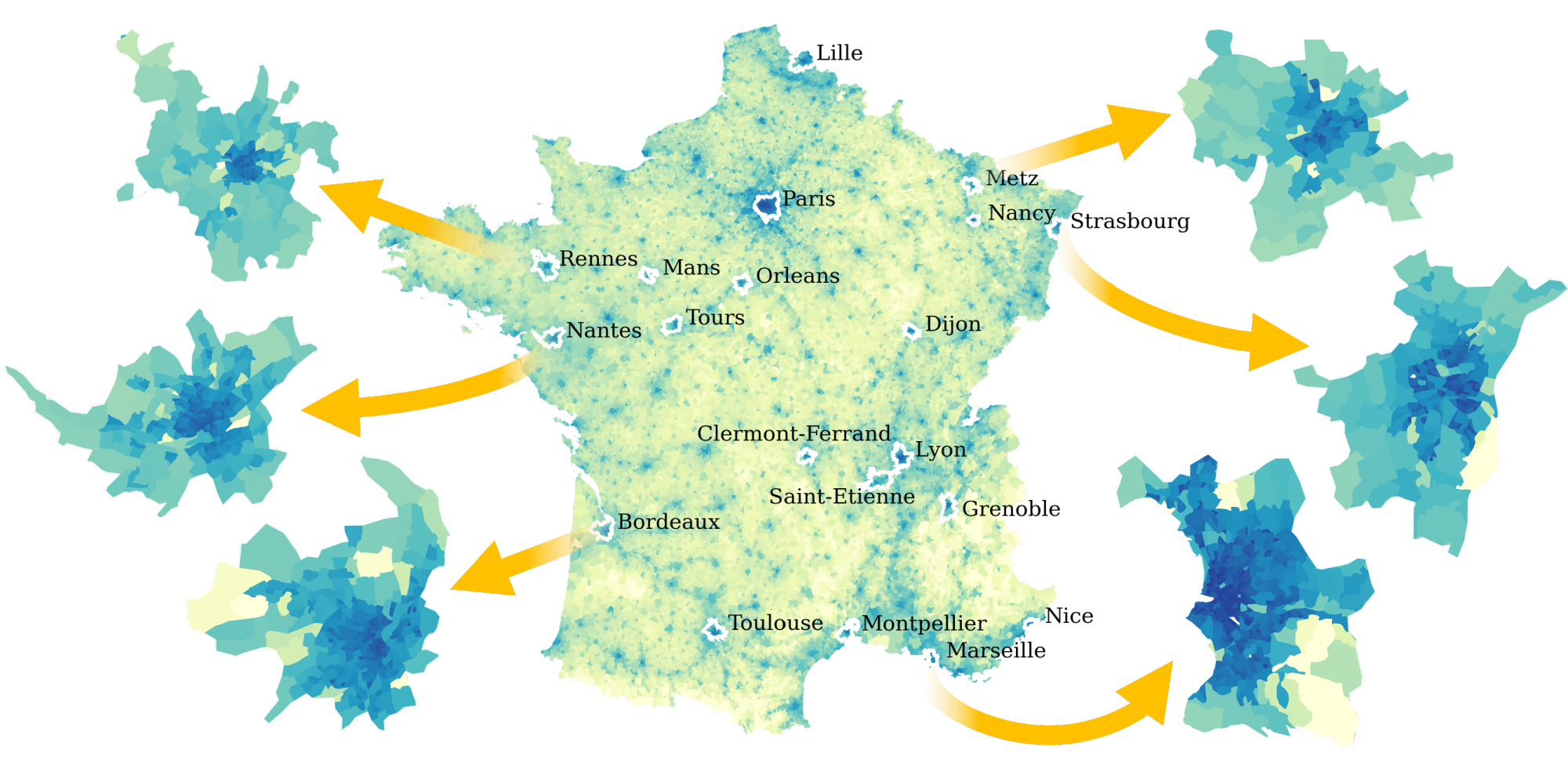}}
\subfloat{
\includegraphics[width=0.075\textwidth]{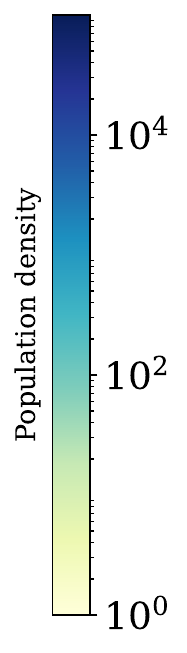}}
\caption{Population density map of France, highlighting the $\bm{20}$ metropolitan areas covered by our dataset. Zoomed-in maps show the heterogeneity of population density in six representative cities, as indicated by the arrows.}
\label{fig:cities}
\end{figure*}

Despite their prospective impact on a wide range of domains, mobile network data are hard to come by. The sensitivity of the information they provide, the concerns for the privacy of the data subjects, and the questions on the advantage that they could provide to market competitors are all reasons why mobile network operators and service providers typically regard the data as confidential, and are not prone to share them with the research community. It limits access to mobile network data, curbing innovation as well as preventing verifiability and reproducibility of the research results whenever permission to use some data is granted under restrictive Non-Disclosure Agreements (NDAs).

In this paper, we present a novel dataset of mobile data traffic that is open to the research community within the context of a challenge organized jointly with the NetMob 2023 conference%
\footnote{NetMob is the primary conference on the analysis of mobile network data to investigate social, urban, societal and industrial problems. Details can be found at \url{https://netmob.org/}.}.
By doing so, we revive a recent tradition of challenges based on mobile phone data that have sparked substantial interest and a flurry of original research outcomes a few years ago. The Data for Development (D4D) challenges organized by Orange in collaboration with NetMob in 2013 and 2014~\cite{d4d_ivory,d4d_senegal}, the International Telecommunication Union (ITU) challenge to investigate the Ebola epidemic in West African countries in 2015~\cite{ITU_Project}, 
the Telecom Italia Big Data Challenge launched in 2014 and 2015~\cite{barlacchi2015multi}, the Data for Refugees (D4R) challenge conducted by Turk Telekom in 2018~\cite{salah18d4r} or the Future Cities Challenge supported by Foursquare during the NetMob 2019 conference~\cite{foursquare18} are prominent examples of such past initiatives.

Our dataset, referred to as \data hereinafter, sets a new standard for mobile network data made available to the research community, from multiple perspectives as follows.
\begin{itemize}[noitemsep,topsep=2pt]
\item While previous datasets have largely focused on Call Detail Records (CDRs) that only capture network events associated with voice calls and text messages that are sparse and irregular over time, \data contains information about the data traffic generated by the mobile devices attached to a modern 4G cellular network, which has been for the past ten years the vastly predominant way of accessing wireless network services.
\item The \data dataset captures traffic in a developed country like France, which offers a different perspective than earlier datasets focusing on developing countries; also, the data spans $20$ metropolitan areas in France, offering the possibility of generalizing analyses and juxtaposing results across heterogeneous urbanization levels and population densities.
\item Unlike any dataset previously available to the research community within the framework of open challenges, \data offers rich information about the usage of $68$ popular mobile services, which opens significant opportunities to understand the consumption of applications and its implications across research domains.
\item The original generation process behind the \data dataset makes a major step beyond the legacy approach of using Voronoi tessellations as a proxy for antenna coverage, and results in a dataset of unprecedented spatial accuracy where the mobile data traffic information is mapped to more than $870,000$ high-resolution regular grids whose individual elements span $100\times100$ m\textsuperscript{2} each, for a total of over $440$ billion data points.
\end{itemize}

Overall, the \data dataset has the potential to support many novel explorations of mobile network traffic and enable the development of new applications on top of those findings. We look forward to seeing creative and constructive uses of the data by the research community worldwide.


\vspace*{-8pt}
\section{Data sources}
\label{sec:data}

The generation process underpinning \data hinges upon open-source geospatial data and extensive measurements from Orange, a major mobile network operator in France. 
We note that Orange roughly has a $30$\% market penetration in the country that is relatively uniform across the French territory. This provides a solid statistical basis for downstream analyses that generalize to the entirety of the local population.
We next present the different data sources, and discuss the ethical standards of the data collection and processing.

\begin{figure*}[tb]
\centering
\includegraphics[width=0.85\textwidth]{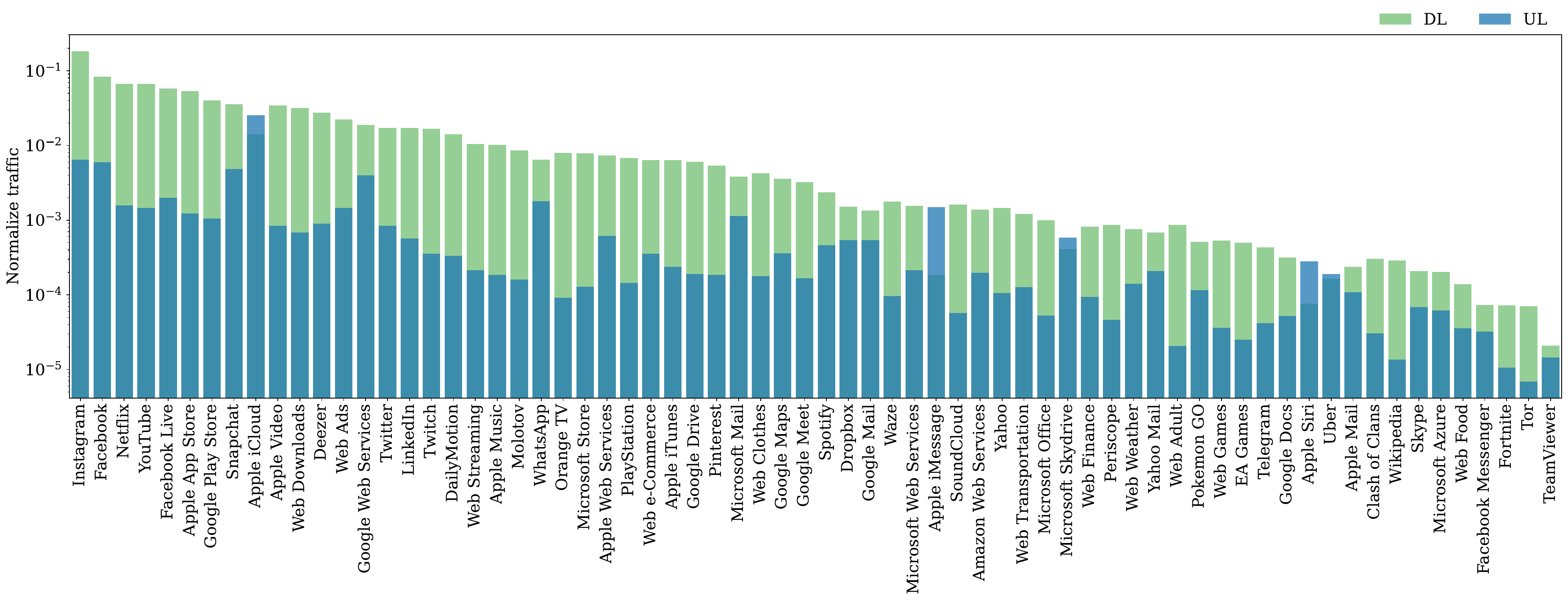}
\vspace{-8pt}
\caption{Mobile services in the \data dataset, ranked by the fraction of total demand they generate.}
\label{fig:apps_traffic}
\vspace{-8pt}
\end{figure*}

\vspace*{-4pt}
\subsection{Metropolitan areas}
\label{sub:data-cities}

We use geographical shapefiles outlining the $20$ major \textit{metropolitan regions} in France, 
a zoning employed by national and local administrations to carry out joint planning of the educational, cultural, economic, and social initiatives over the country territory.
Each metropolitan region encompasses a set of neighboring \textit{communes}, which are French local administrative zones analogous to a civil township in the United States, and thus covers both dense urban, suburban, and more rural areas that constitute the conurbation of a specific city.
Figure~\ref{fig:cities} shows the location and contour of all metropolitan areas covered by the \data dataset, overlapped to a country-wide map of population density. The areas of a few sample cities are magnified to appreciate the diversity of population density captured by each of them.

Overall, our dataset includes the foremost industrial, commercial, and financial centers of France, which are home to more than one third of the total population in the country. 

\vspace*{-4pt}
\subsection{Mobile network traffic}
\label{sub:data-traffic}

We employ mobile data traffic collected over the $20$ target metropolitan areas for $77$ consecutive days, \ie roughly two and a half months, from March $16$, $2019$, to May $31$, $2019$. 

\subsubsection{Service-level traffic volumes.}
The traffic measurements were performed by Orange using passive measurement probes tapping at the Gi, SGi and Gn interfaces connecting the Gateway GPRS Support Nodes (GGSNs) and the Packet Data Network Gateways (PGWs) of the of Long Term Evolution (LTE) Evolved Packet Core (EPC) network to external public data networks (PDNs). This monitoring strategy allows capturing all 4G traffic traversing the mobile network serving the whole country.
The probes run dedicated proprietary classifiers that allow associating individual traffic flows to their corresponding mobile applications to enable network monitoring, traffic engineering and research activities.

Ultimately, the EPC probes gather information about the uplink (UL) and downlink (DL) volume of the demand generated by each of $68$ mobile services, which are together responsible for about $70$\% of the total mobile data traffic observed by Orange in France.
Figure \ref{fig:apps_traffic} offers a look into the applications included in the dataset, and a ranking of the same based on the fraction of total traffic they generate. We observe that the services responsible for the largest demands are responsible for around $10$\% of the overall mobile network usage. The logarithmic scale of the ordinate highlights the power law that characterizes the relative consumption across applications, and the consequent diversity of per-service traffic, which spans three orders of magnitude when juxtaposing the top and bottom applications in the dataset.

\subsubsection{Traffic flow to eNodeB association.}
We associate traffic volumes to specific base stations using network signalling data (NSD) captured by probes monitoring the LTE S1 interface connecting eNodeBs, \ie 4G base stations, to the Mobility Management Entity (MME). 
The NSD allow tracking the eNodeB serving a mobile device across all control-plane events, which include ($i$) voice and texting communications such as call establishments and SMS transmissions, ($ii$) handovers, \ie device cell changes during communication, ($iii$) Tracking Area (TA) updates, \ie cell changes that cross boundaries among larger regions that trigger control messages also from idle devices), ($iv$) active paging, \ie periodic requests to update the location of the device started from the network side, ($v$) network attaches and detaches generated by devices joining or leaving the network as they are turned on/off, or ($vi$) data connections, \ie requests to assign resources for traffic generated by mobile applications running on the device.
Such high-frequency NSD events allow associating each traffic flow to the exact sequence of eNodeBs that serviced it, and thus to accurately link (portions of) the traffic volume of the flow to each base station.

\subsubsection{Service-level traffic time series at eNodeBs}

We aggregate the traffic volume generated by all uplink or downlink flows pertaining to a given mobile service and served by a same eNodeB.
We also aggregate the resulting traffic volumes over $15$-minute time intervals, \ie a temporal granularity that allows observing a wide range of time-dependent phenomena, while keeping the dataset at a reasonable size.

\begin{figure*}
\centering
\includegraphics[width=.95\textwidth]{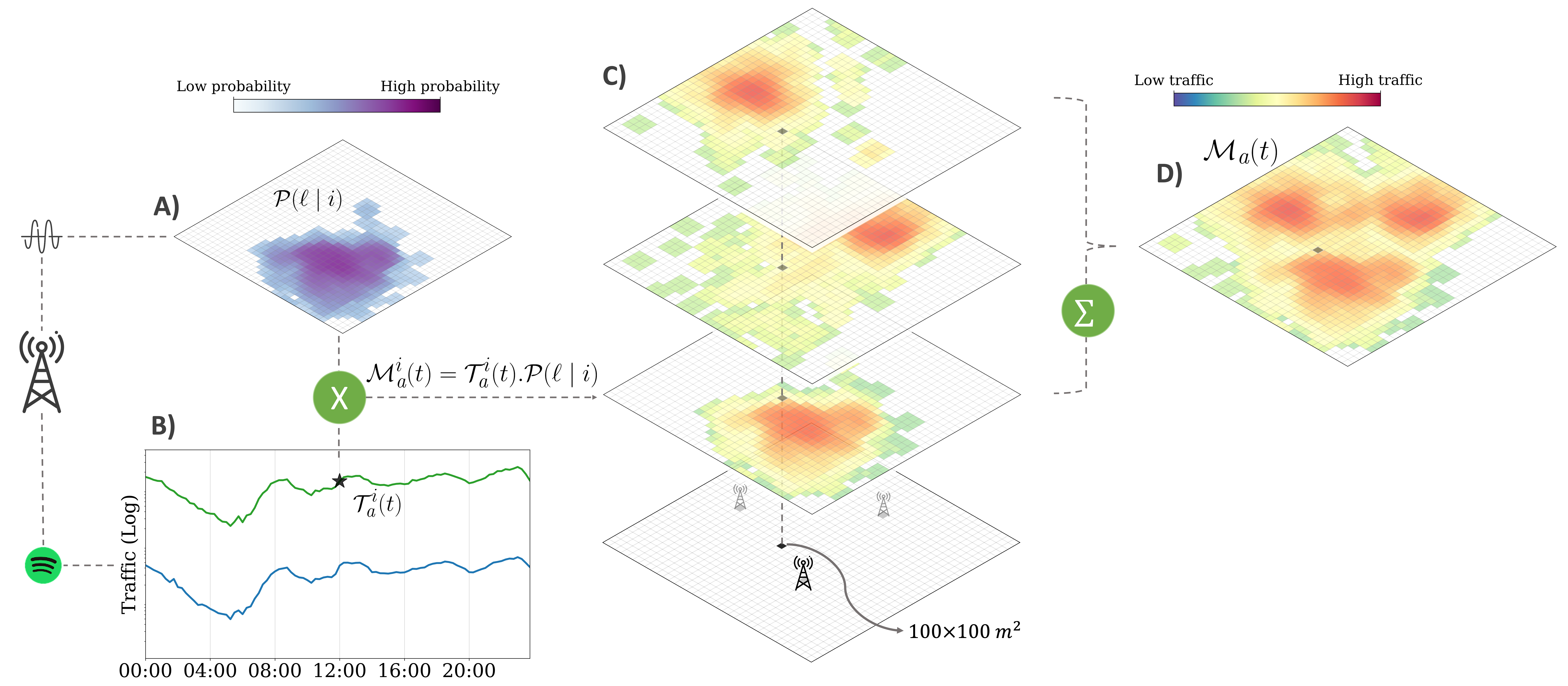}
\caption{Summary of the dataset generation methodology. (A) Coverage matrix for eNodeB $\bm{i}$. (B) Traffic time series aggregated over all users for application $\bm{a}$, \ie Spotify, at eNodeB $\bm{i}$. (C) Spatial mapping of the Spotify traffic at eNodeB $\bm{i}$ during time step $\bm{t}$. (D) Overall spatial distribution of Spotify traffic in $\bm{t}$, as the sum of all eNodeB maps.}
\label{fig:algorithm}
\vspace*{-4pt}
\end{figure*}

Overall, the mobile network traffic data we employ is in the form of time series of the uplink and downlink load generated by $68$ mobile applications at every eNodeB, with $15$-minute time steps.
In order not to disclose the sensitive information of the actual volume of traffic served by the mobile network operator, we normalize all traffic by a same random value. Therefore, traffic is not provided with a specific unit (\eg bytes) but it is still fully comparable across space and time.

\vspace*{-4pt}
\subsection{Coverage dataset}
\label{sub:data-coverage}

Having associated the correct service-level traffic to each eNodeB over time, we employ coverage information for each eNodeB to perform a spatial mapping of the traffic information to the geographical space.
This allows producing data with a much higher spatial resolution than using legacy approaches such as Voronoi tessellations.

Specifically, we start from coverage data computed using a commercial radio-frequency signal propagation tool. For every eNodeB, coverage is encoded as probabilities of association over a regular grid tessellation with tiles of $100\times100$ m$^2$ each. A bi-dimensional matrix of $600\times600$ tiles is produced for a single eNodeB, hence providing complete coverage information over an area of $60$ km$^2$ surrounding the base station.
The matrix tiles contain probabilities $p(i \mid\ell)$ that explain the likelihood of a User Equipment (UE) to connect to eNodeB $i$ while at tile $\ell$. By inverting this probability information via Bayes' theorem, under the assumption of a uniform density of population within the area covered by the eNodeB, we derive $p(\ell \mid i)$, \ie the probability that a device associated with a base station $i$ is located at tile $\ell$. We use $p(\ell \mid i)$ to distribute over the high-resolution grid of $100\times100$-m$^2$ tiles the service-level traffic observed by each eNodeB in a probabilistic fashion, as detailed in Section~\ref{sec:method}.

\vspace*{-4pt}
\section{Generation methodology}
\label{sec:method}

The different data presented above are combined to generate the final \data dataset, following a process that is visually summarized in Figure~\ref{fig:algorithm}.
Formally, let us denote by $\mathcal{T}_{\textit{a}}^{i}(t)$ the mobile traffic generated by application $a$ at eNodeB $i\in I$ during time slot $t\in T$, where $I$ is the set of all base stations and $T$ denotes the whole system observation period.
Also, recall that $\mathcal{P}(\ell\mid i)$ is the probability of a user to be at a location $\ell$ while being connected to eNodeB $i$.

As portrayed in Figure~\ref{fig:algorithm}, we first multiply traffic observed at each eNodeB (denoted by B in the figure) by the UE positioning probability of the same eNodeB (A in the figure), so as to distribute over space the traffic volume observed at the base station.
This operation is repeated for each time slot $t$ of the time series of every mobile service (\eg Spotify in the figure).
The result is a traffic map $\mathcal{M}_{\textit{a}}^{i}(t)$, which represents how the traffic generated by a given service $a$ (again, Spotify in the figure example) at the target eNodeB $i$ is probabilistically distributed over the geographical space at time $t$.

\begin{table*}[tb]
\footnotesize
\begin{center}
\setlength{\tabcolsep}{4pt}
\renewcommand{\arraystretch}{.34}
\begin{tabular}
{>{\columncolor{gray!20}}c c >{\columncolor{gray!20}}c c >{\columncolor{gray!20}}c c >{\columncolor{gray!20}}c c >{\columncolor{gray!20}}c c >{\columncolor{gray!20}}c c >{\columncolor{gray!20}}c c >{\columncolor{gray!20}}c c >{\columncolor{gray!20}}c c >{\columncolor{gray!20}}c c >{\columncolor{gray!20}}c c}

Region& 
\rotate{Bordeaux}  & 
\rotate{Clermont-Ferrand} &
\rotate{Dijon} & \rotate{Grenoble} & \rotate{Lille} & \rotate{Lyon} & \rotate{Mans} & \rotate{Marseille} &  \rotate{Metz} & \rotate{Montpellier} & \rotate{Nancy} & \rotate{Nantes} & \rotate{Nice} & \rotate{Orleans} & \rotate{Paris} & \rotate{Rennes} & \rotate{Saint-Etienne} & \rotate{Strasbourg} &  \rotate{Toulouse} & \rotate{Tours} & 
 \\
 \hline
 \noalign{\vskip .3ex}
 Rows & 334 & 208 &  195 & 409 & 330 & 426 & 228 & 211 & 226 & 334 & 151 & 277 & 150 & 282 & 409 & 423 & 305 & 296 & 280 & 251  \\
\noalign{\vskip .1ex}

 Columns &
342 & 268 &  234 & 251 & 342 & 287 & 246 & 210 & 269 & 327 & 165 & 425 & 214 & 256 & 346 & 370 & 501 & 258 & 347 & 270  \\
\noalign{\vskip .3ex}
  \hline

\end{tabular}
\vspace*{4pt}
\caption{Dimensions of the matrices of regular grid tiles in each of the $\bm{20}$ target urban regions.}
\label{tab:rows_column_for_cities}
\end{center}
\vspace*{-16pt}
\end{table*}

Since the same regular grid tessellation is used for probabilities $\mathcal{P}(\ell\mid i)$ across all base stations $i\in I$, it is possible to repeat the process above for all eNodeBs, and generate consistent maps of the spatial traffic of a same application $a$ over the whole mobile network.
This is attained by simply adding the different traffic maps as $\mathcal{M}_{\text{a}}(t) =  \sum_{i=1}^{I} \mathcal{M}_{\text{a}}^{i}(t)$ (shown in Figure~\ref{fig:algorithm} as C).
The final $\mathcal{M}_{\text{a}}(t)$ is the overall traffic map of service $a$ at time $t$, defined over locations $\ell$ (D in the figure).

We repeat the above steps for all $t\in T$ and for all applications to ultimately obtain a complete spatiotemporal representation of the service-level traffic in the $20$ target metropolitan areas introduced in Section~\ref{sub:data-cities} over a high-resolution grid of locations $\ell$.

\vspace*{-4pt}
\subsection{Ethics considerations}
\label{sec:data-ethics}

The mobile network traffic dataset we use to generate the dataset was collected, processed and aggregated as described in Section~\ref{sub:data-traffic} in full compliance with Article 89 of the General Data Protection Regulation (GDPR)~\cite{gdpr}, under the supervision of the Data Protection Officer (DPO) at Orange. In particular, all data management was performed on a secure platform at the operator's premises and the raw data was deleted immediately afterward.

The resulting service-level time series represent traffic aggregated over all UEs both in space, at eNodeB level, and time, over $15$-minute intervals. Moreover, the traffic associated to different base stations is further aggregated via the spatial mapping described earlier. The final representation does not allow re-identifying or tracking individual users.

\vspace*{-4pt}
\subsection{Final dataset format}

To facilitate access to the data, the \data dataset is divided into spatial representations and traffic information, which are detailed next. As part of the material provided with the dataset, we also make available Jupyter notebooks in Python with examples of manipulations of the dataset and calculations of high-level statistical indicators and plots%
\footnote{\url{https://github.com/nds-group/netmob2023challenge}.}.

\subsubsection{Spatial representation}
We publish a GeoJson file for each metropolitan region with the WGS84 coordinate system. Each of these files contains the \texttt{tile} identifier of each regular grid tile in the target urban area, and the corresponding polygon that bounds the $100 \times 100$-m$^2$ geographical surface of the tile. We also provide an alternative matrix representation of the space, in the form of a text file containing the number of rows and columns in the matrix, whose values are also shown in Table~\ref{tab:rows_column_for_cities}.

\begin{figure}
\centering
\includegraphics[width=\columnwidth]{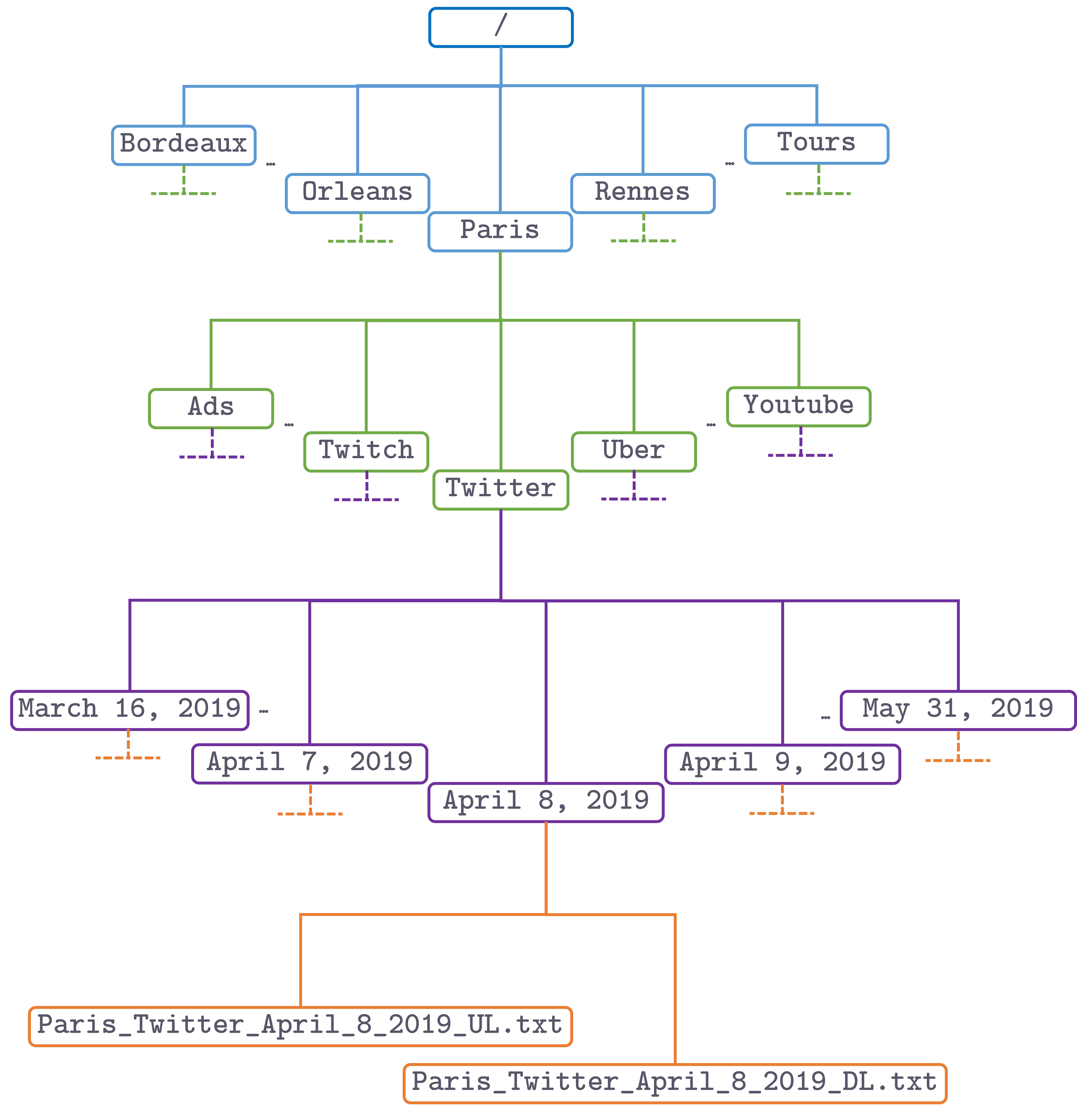}
\caption{Hierarchical organization of traffic files.
\label{fig:folder_tree}}
\end{figure}

\subsubsection{Traffic information}
The traffic dataset folder is organized according to a tree structure, as shown in Fig~\ref{fig:folder_tree}. The root is the target urban region, followed by the application name and then the day, with two leaves, \ie text files, for the UL and DL directions, respectively. 
In each text file, one line represents the (normalized) traffic in a given tile of the city, for the corresponding application and date, from midnight to \texttt{23:45} at 15-minute time steps, as shown in Table \ref{tab:traffic_records}.

More precisely, each line contains the following fields:
\begin{itemize}
    \item \texttt{tile} is the tile identifier used in the GeoJson spatial representation file for the same urban area. In the case of the matrix representation of tiles, the position inside the matrix can be retrieved by an integer division and a modulo operation.
    \item \texttt{t\textsubscript{hh:mm}} is a vector of 96 values, corresponding to the traffic observed in UL or DL at location tile for the target city, service, and day, during all 15-minute time slots, starting from midnight to $23$:$45$. Due to daylight saving time in France on March 31, 2019, between $02$:$00$ and $03$:$00$, these traffic records have $92$ columns. Therefore, the first eight columns cover the period from $00$:$00$ to $01$:$45$, and the ninth column corresponds to $03$:$00$.
\end{itemize}

\begin{table}[tb]
\begin{center}
\begin{tabular}{c c c c c}
    \texttt{tile} & \texttt{t\textsubscript{00:00}} & \texttt{t\textsubscript{00:15}} & \dots & \texttt{t\textsubscript{23:45}} \\
    \hline
    \dots & & & & \\
    1966 & $\mathcal{T}_0$ & $\mathcal{T}_1$ & \dots & $\mathcal{T}_{95}$\\
    1967 & $\mathcal{T}_0$ & $\mathcal{T}_1$ & \dots & $\mathcal{T}_{95}$\\
    1968 & $\mathcal{T}_0$ & $\mathcal{T}_1$ & \dots & $\mathcal{T}_{95}$\\
    \dots & & & & \\
\end{tabular}
\vspace*{4pt}
\caption{Example of the format of a traffic file.}
\label{tab:traffic_records}
\end{center}
\vspace*{-20pt}
\end{table}

\section{Qualitative analysis}
\label{sec:validation}

In this section, we provide a preliminary exploration of some characteristics of the \data dataset. The analysis aims at discussing basic properties of the spatial and temporal evolution of the service-level mobile traffic, and highlight anomalies inherent to the dataset that may bias downstream usages of the same.

\subsection{Anomaly analysis}

\begin{figure}
\centering
\subfloat[Bordeaux, Facebook Live]
{\includegraphics[width=0.95\columnwidth]{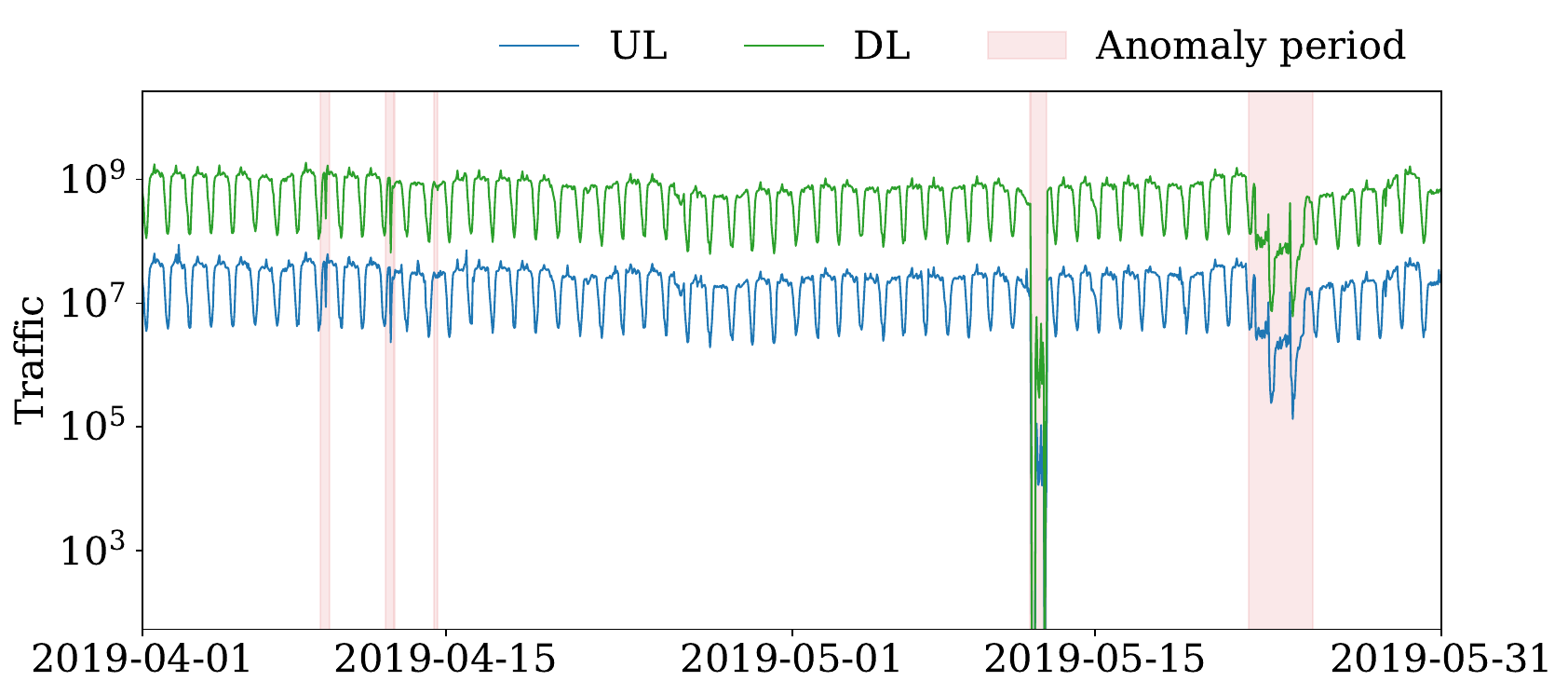}}\\
\subfloat[Paris, Instagram]{
\includegraphics[width=0.95\columnwidth,trim={0 0 0 35pt},clip]{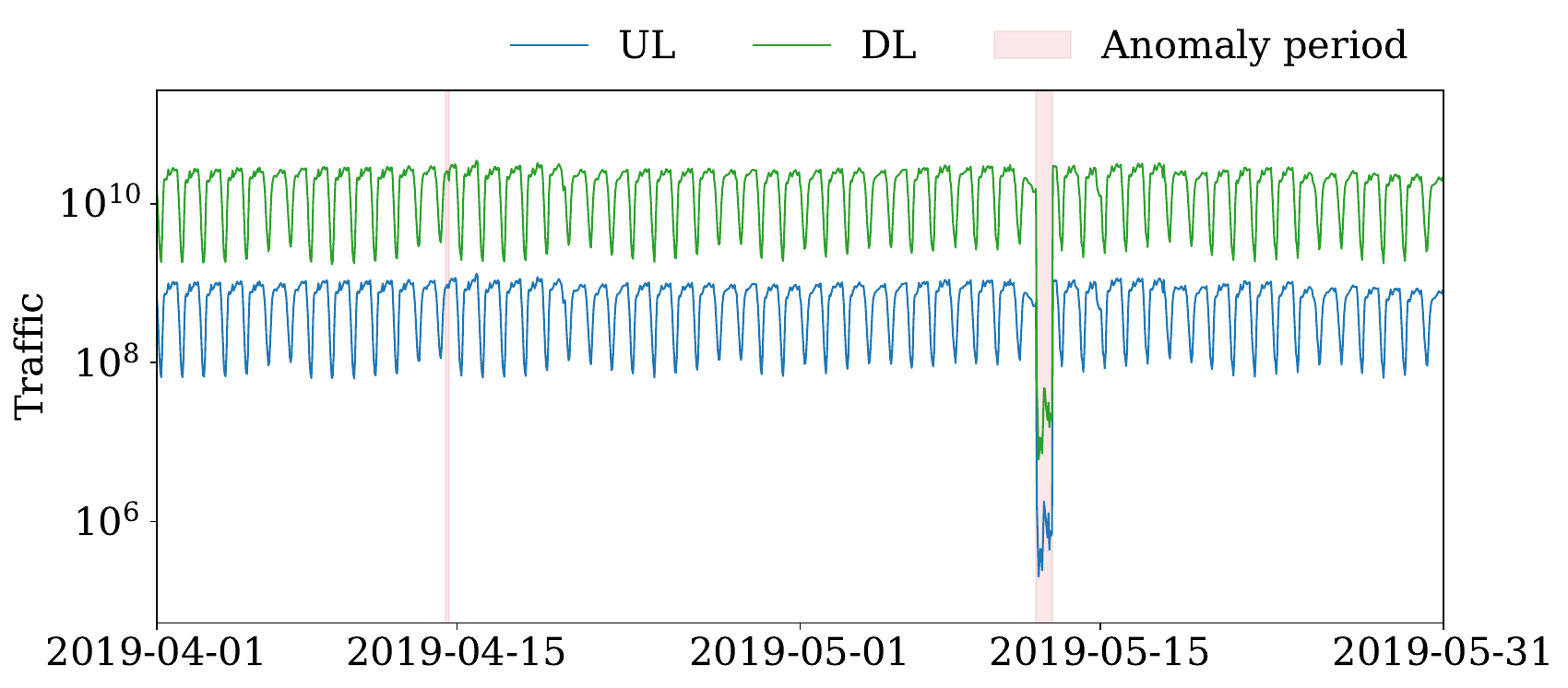}}\\
\subfloat[Toulouse, Twitter]{\includegraphics[width=0.95\columnwidth,trim={0 0 0 35pt},clip]{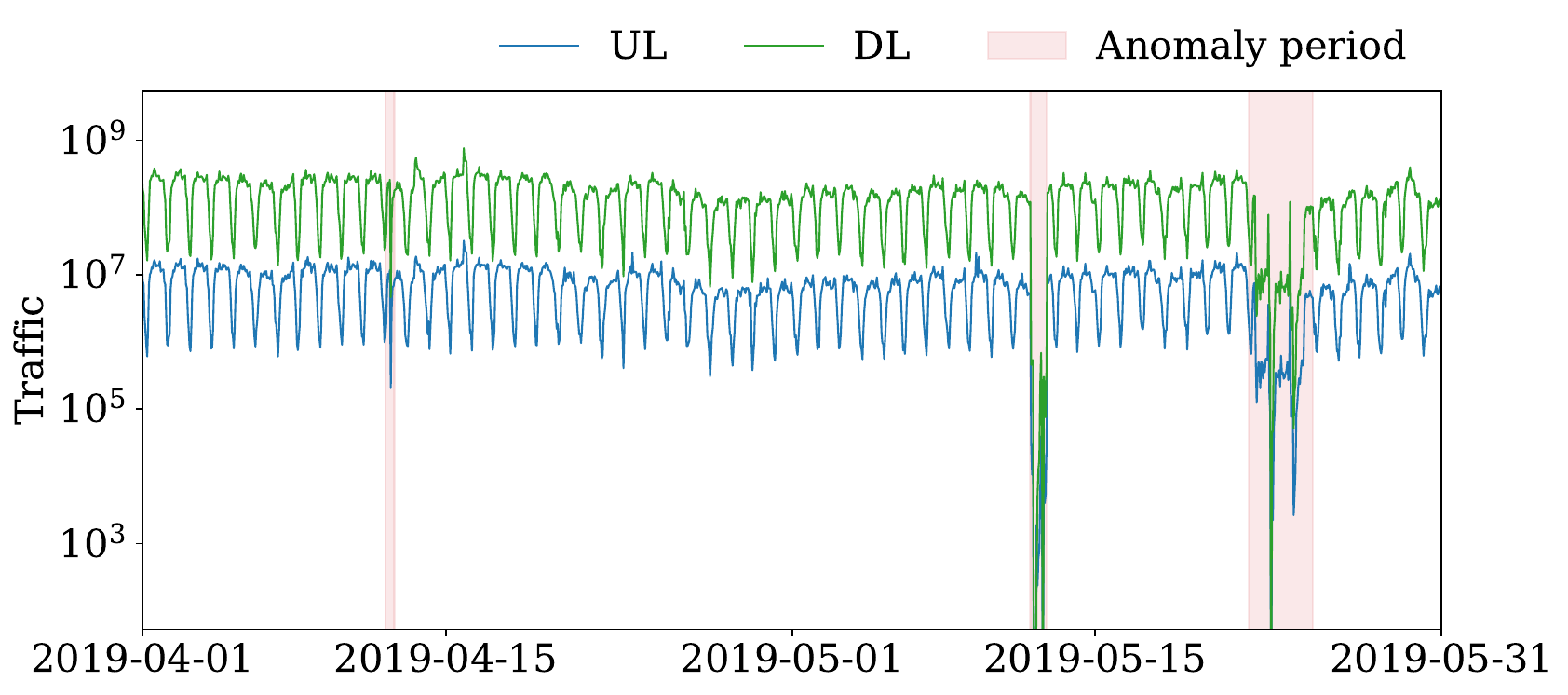}}
\vspace{-8pt}
\caption{Traffic time series highlighting different anomalies in various cities of the \data dataset.
\label{fig:anomaly_plot}}
\vspace*{-4pt}
\end{figure}

The \data dataset is based on traffic information measured in a production network. As such, it is characterized by anomalous events of technical nature, caused, \eg by exceptional surges in demands, radio access network malfunctioning, network configuration errors, energy supply problems, or issues in the traffic monitoring system.  

\begin{table}[tb]
\begin{center}
\begin{tabular}{l l l}
    Period & Affected regions & Services\\
    \hline
    April 9, 2019 & Bordeaux & All\\
    April 12, 2019 & Bordeaux, Toulouse & All\\
    April 14, 2019 \cite{fb_outage_14_4} & Nationwide & Meta Inc.  \\
    May 12, 2019 \cite{china_telcom_wordwide_may_13} &  Nationwide & All \\
    May 22-25, 2019 & Bordeaux, Toulouse & All\\
\end{tabular}
\vspace*{4pt}
\caption{Major anomalies detected in the dataset.}
\label{tab:anomaly_periods}
\end{center}
\vspace*{-24pt}
\end{table}

We report examples of anomalous events that can be observed in the \data dataset in Figure \ref{fig:anomaly_plot}, for different combinations of cities and services.
The plots show the overall traffic in UL and DL, with anomalies evidenced by light red areas.
The three plots show for instance the effects of a nationwide network outage on May 12, 2019, which caused both UL and DL traffic to drop substantially in the whole Orange network.
The top and bottom plots also highlight a an event affecting only the South-East of France but for a longer period spanning from May 22 through May 25: the problem caused a reduction of traffic in Bordeaux and Toulouse, but not in Paris, due to the geographical locations of the cities.
More rarely, outages concern specific applications, such as for the services of Meta on April 14, as seen, \eg in the top plot for Facebook Live traffic.
The full list of major anomalies we detected in the data is provided in Table~\ref{tab:anomaly_periods}.

\subsection{Temporal Analysis}

\begin{figure}
\centering
\subfloat[Paris, Netflix]{\includegraphics[width=.95\columnwidth,trim={0 0 0 0pt},clip]{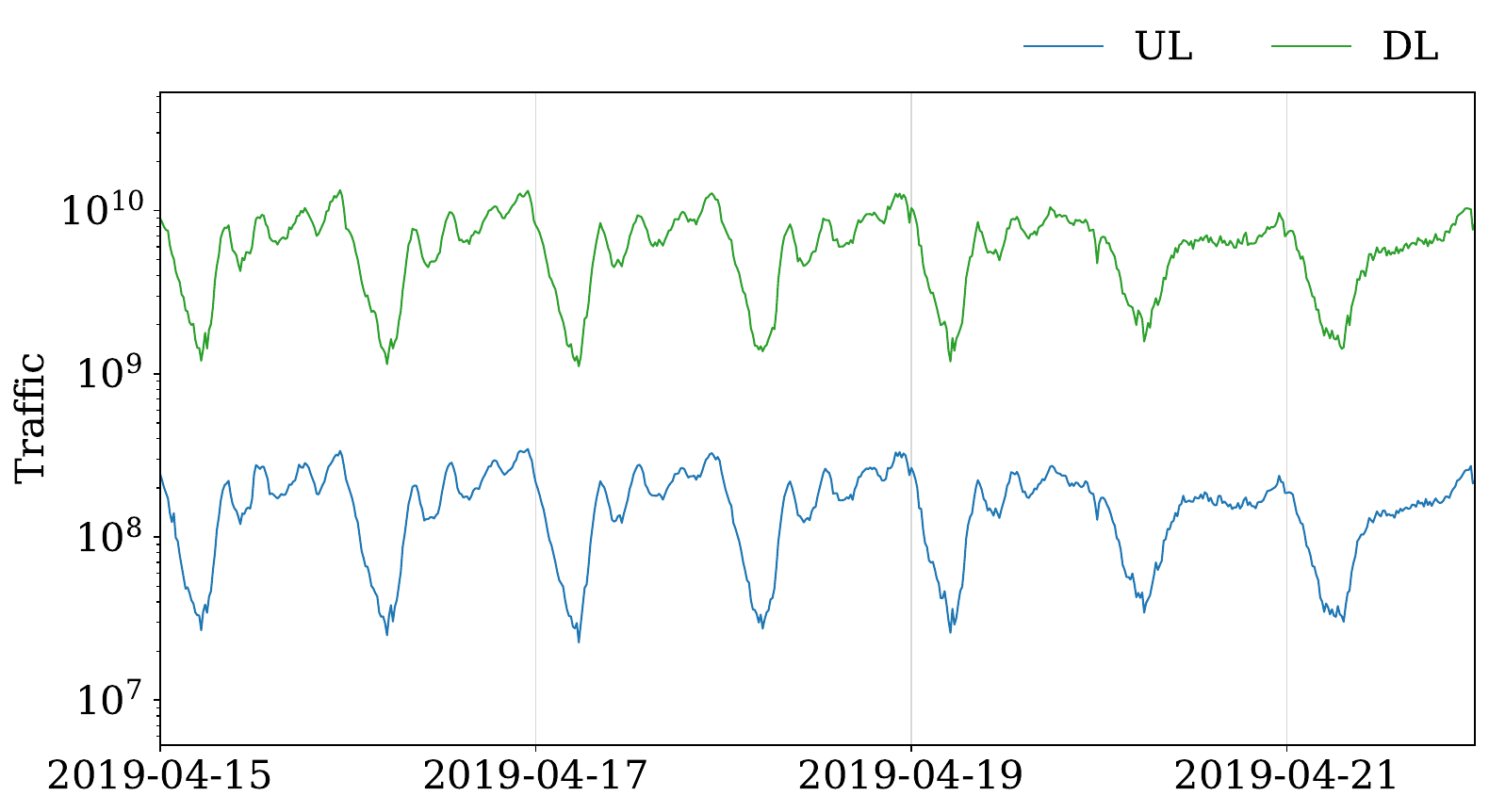}}\\
\subfloat[Paris, LinkedIn]{
\includegraphics[width=.95\columnwidth,trim={0 0 0 35pt},clip]{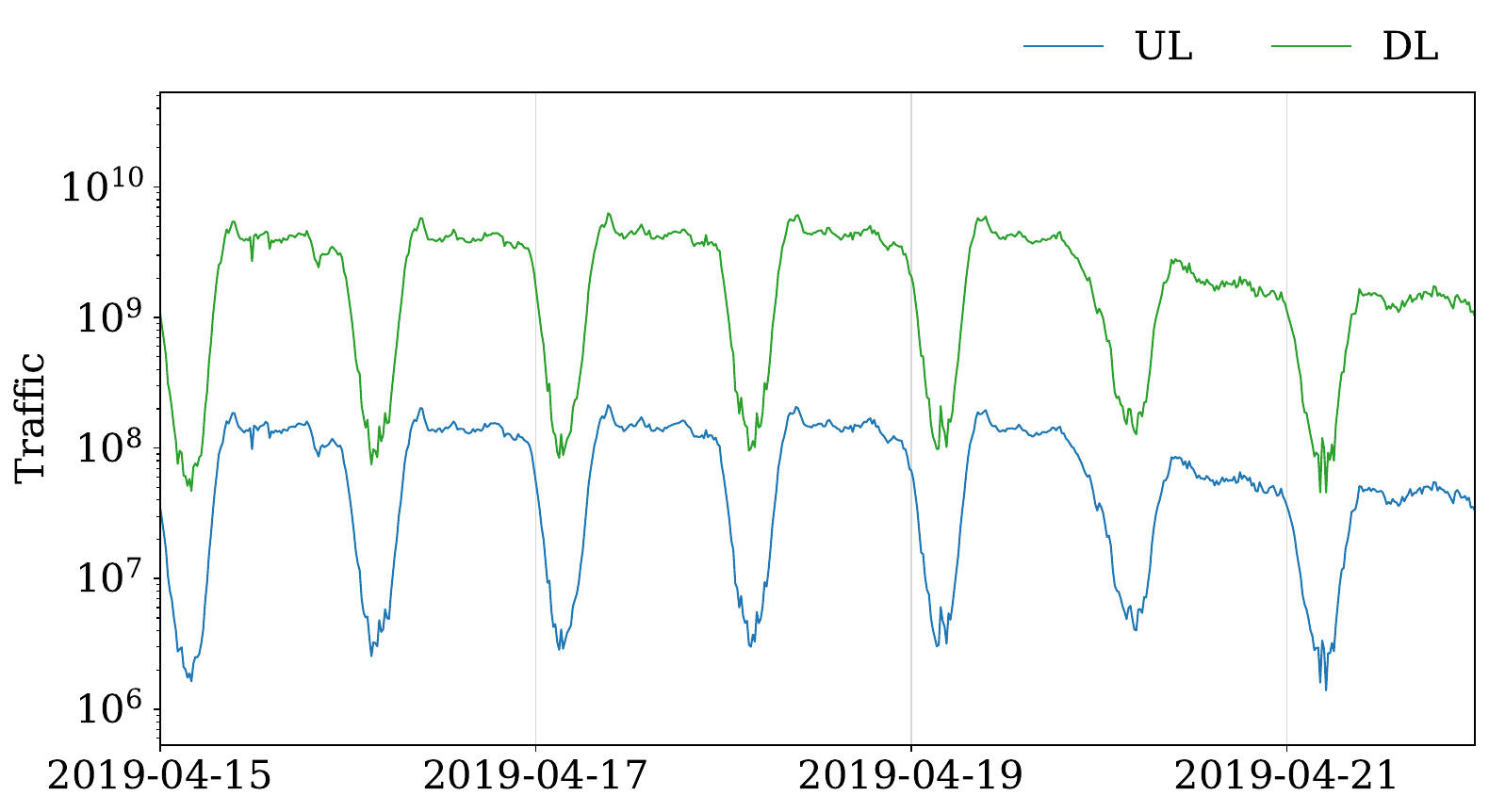}}\\
\subfloat[Paris, Apple iCloud]{\includegraphics[width=.95\columnwidth,trim={0 0 0 35pt},clip]{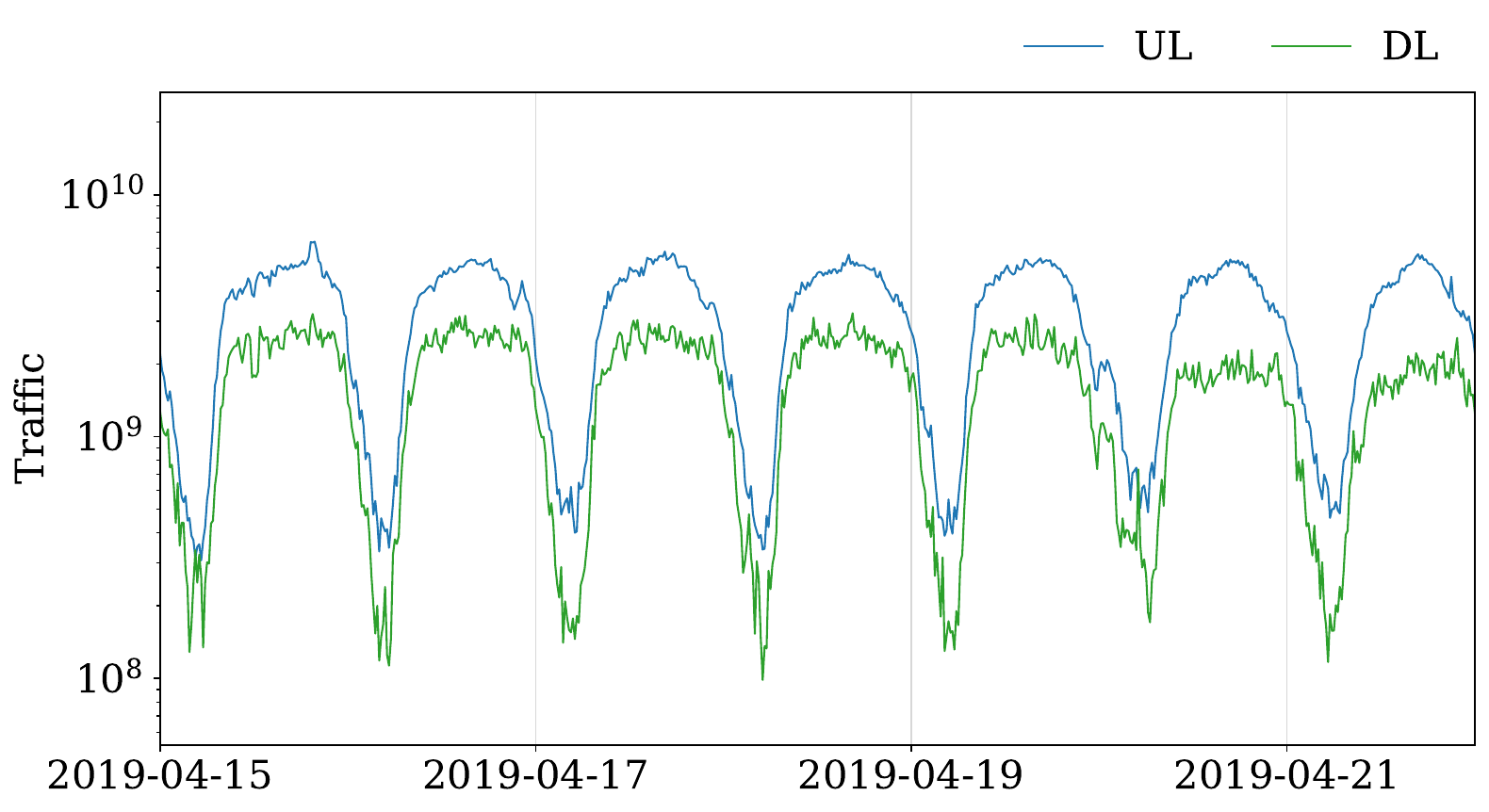}}\\
\subfloat[Paris, Uber]{\includegraphics[width=.95\columnwidth,trim={0 0 0 35pt},clip]{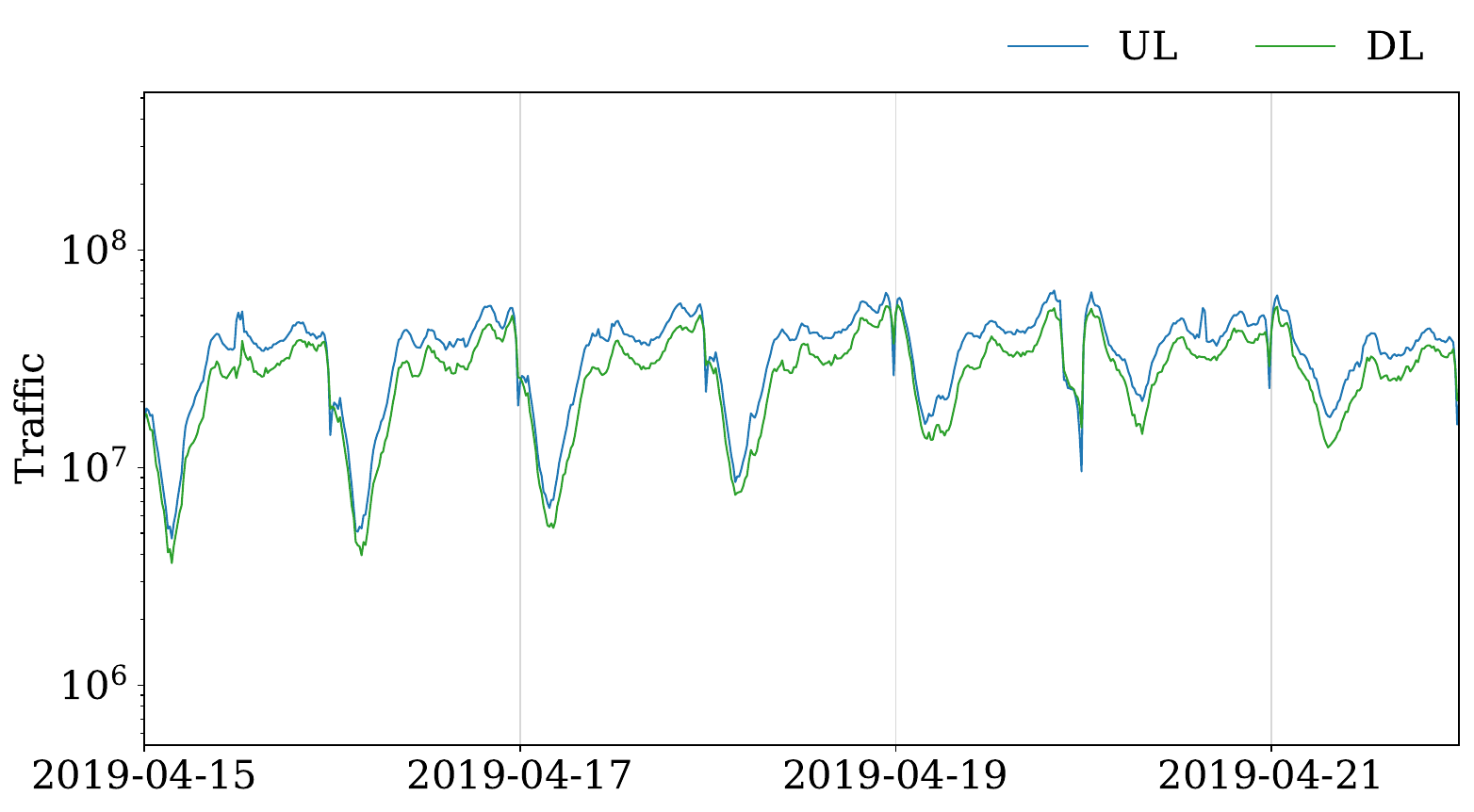}}
\caption{Traffic time series of four different applications in Paris, as observed during a same week.}
\label{fig:app_behaviour_in_city}
\end{figure}

\begin{figure}
\centering
\subfloat[Bordeaux, Netflix]{\includegraphics[width=.945\columnwidth,trim={0 0 0 0pt},clip]{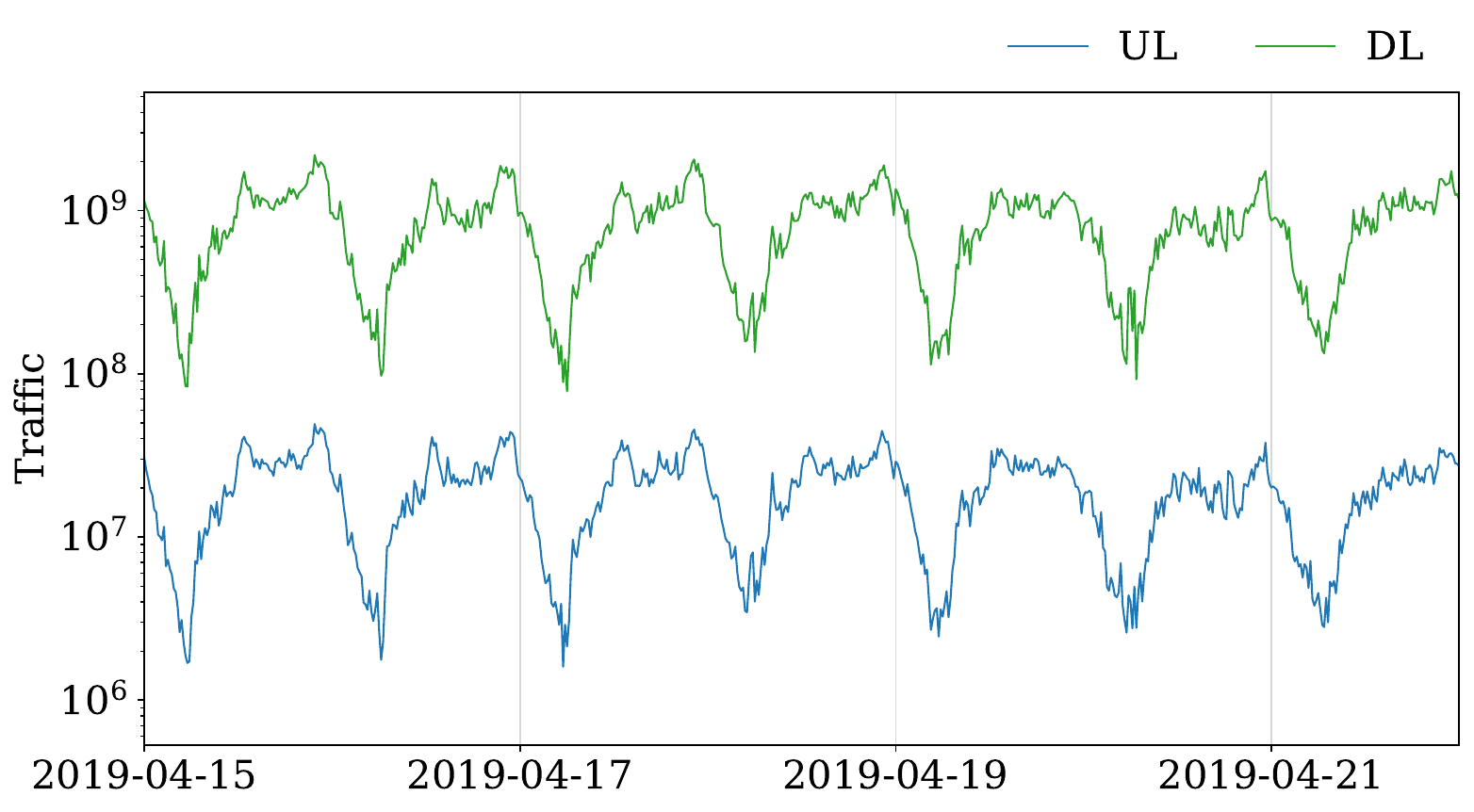}}\\
\subfloat[Lyon, Netflix]{
\includegraphics[width=.945\columnwidth,trim={0 0 0 35pt},clip]{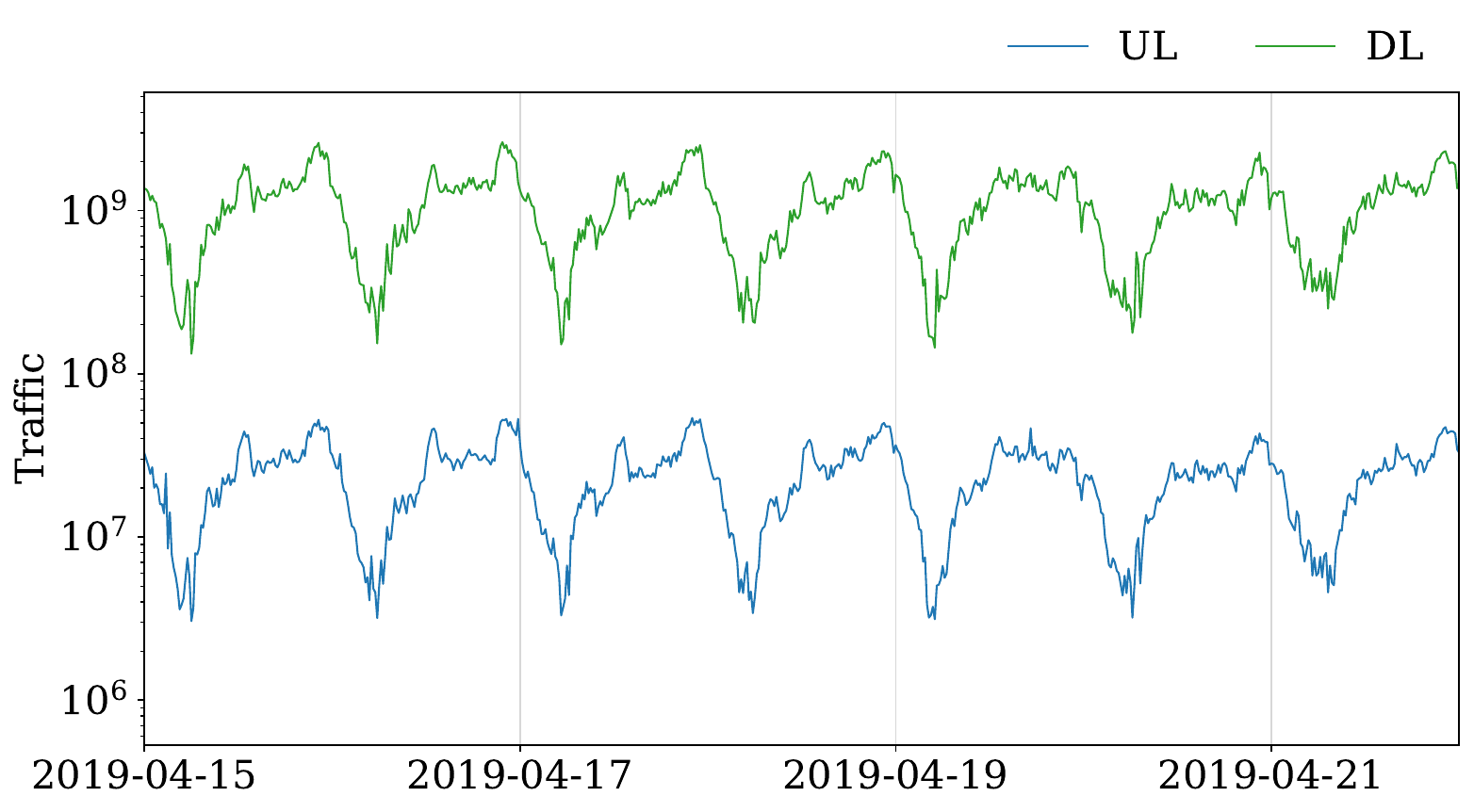}}\\
\subfloat[Toulouse, Netflix]{\includegraphics[width=.945\columnwidth,trim={0 0 0 35pt},clip]{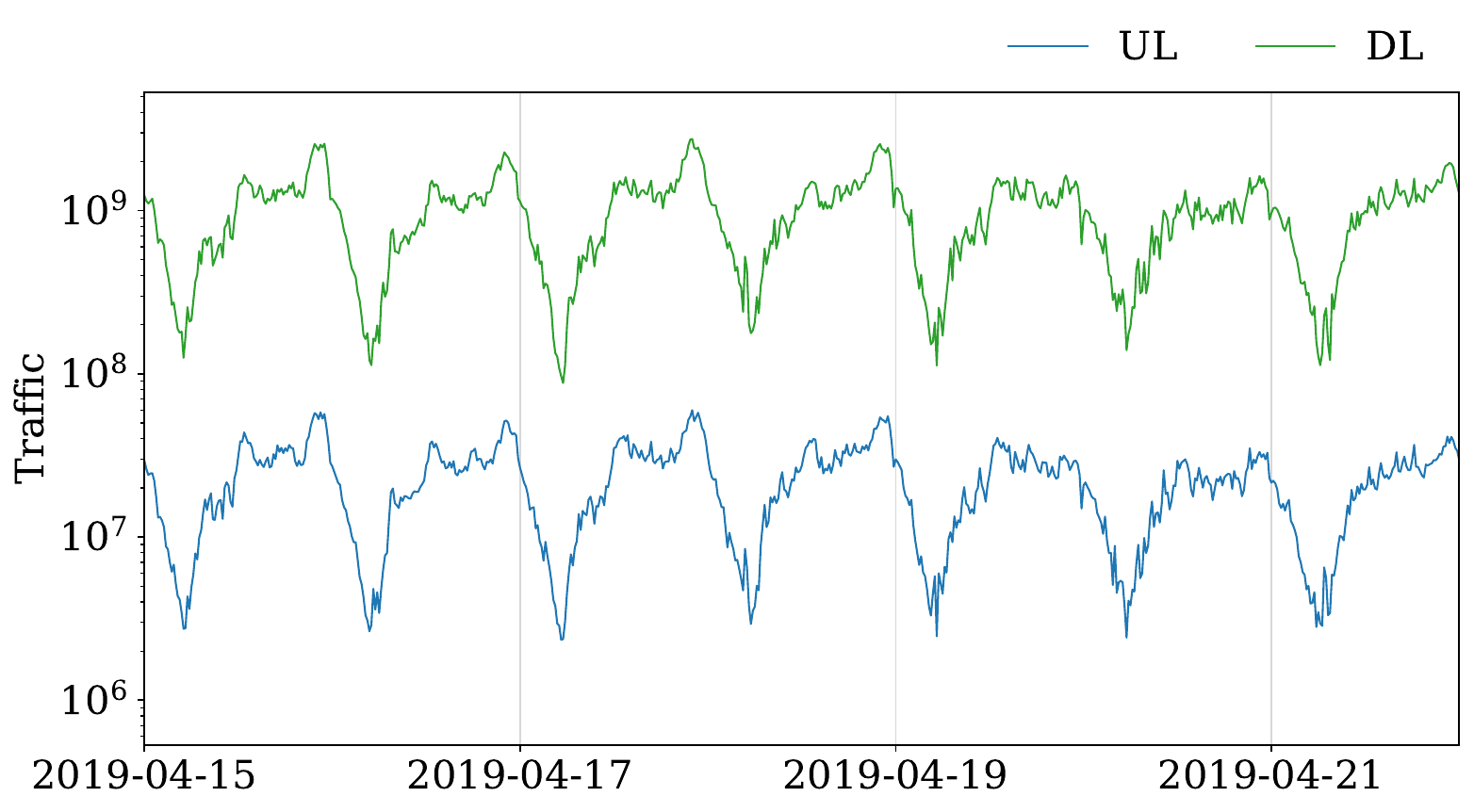}}\\
\subfloat[Marseille, Netflix]{\includegraphics[width=.945\columnwidth,trim={0 0 0 35pt},clip]{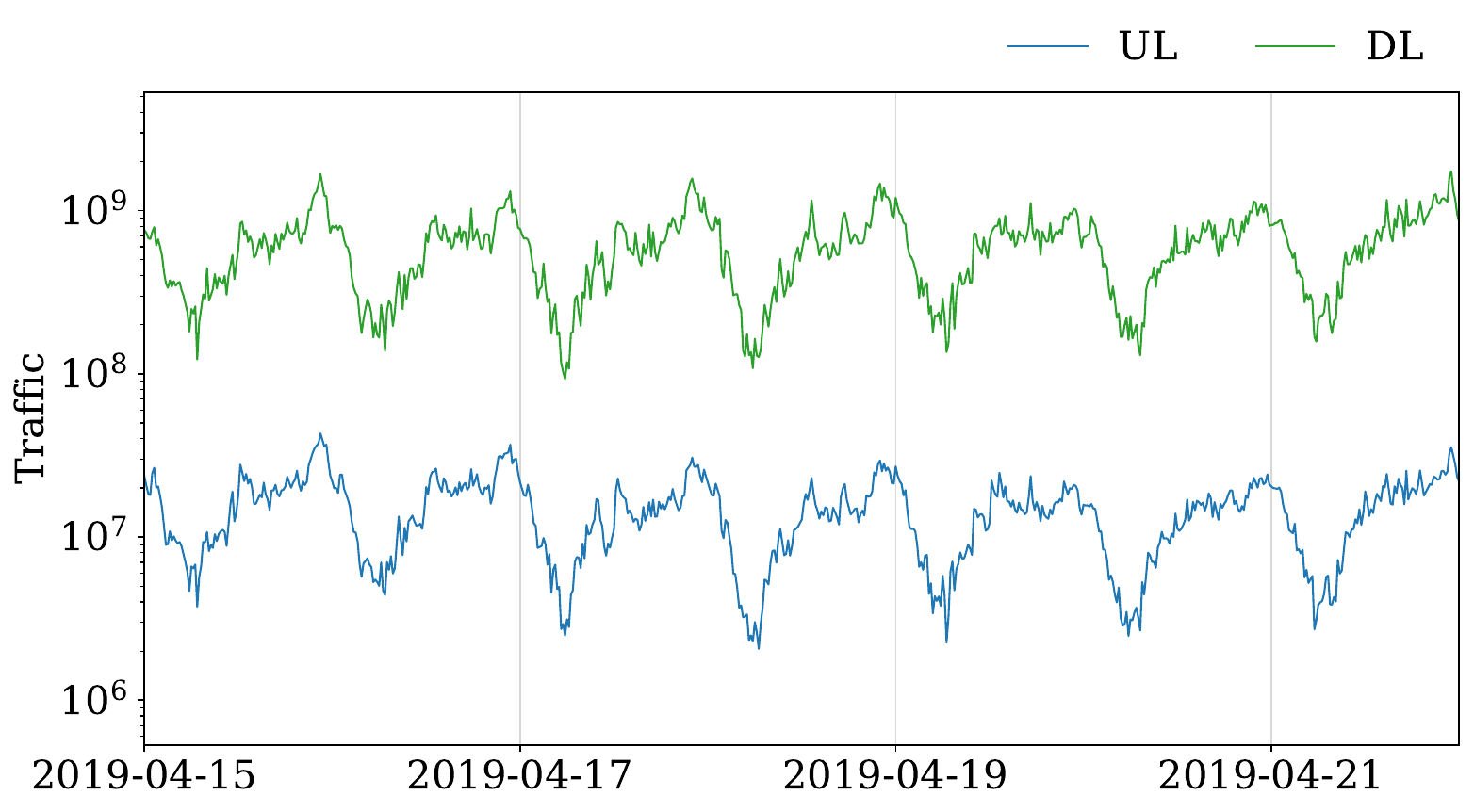}}
\caption{Traffic time series of the Netflix service in four different cities during a same week.}
\label{fig:app_behaviour_in_cities}
\end{figure}

\begin{figure*}
\RaggedLeft
\subfloat{\includegraphics[width=0.2\textwidth]{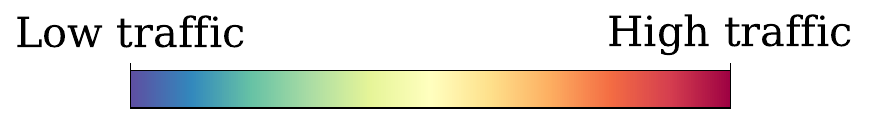}}
\vspace{5pt}
\\
\subfloat{\includegraphics[width=0.24\textwidth]{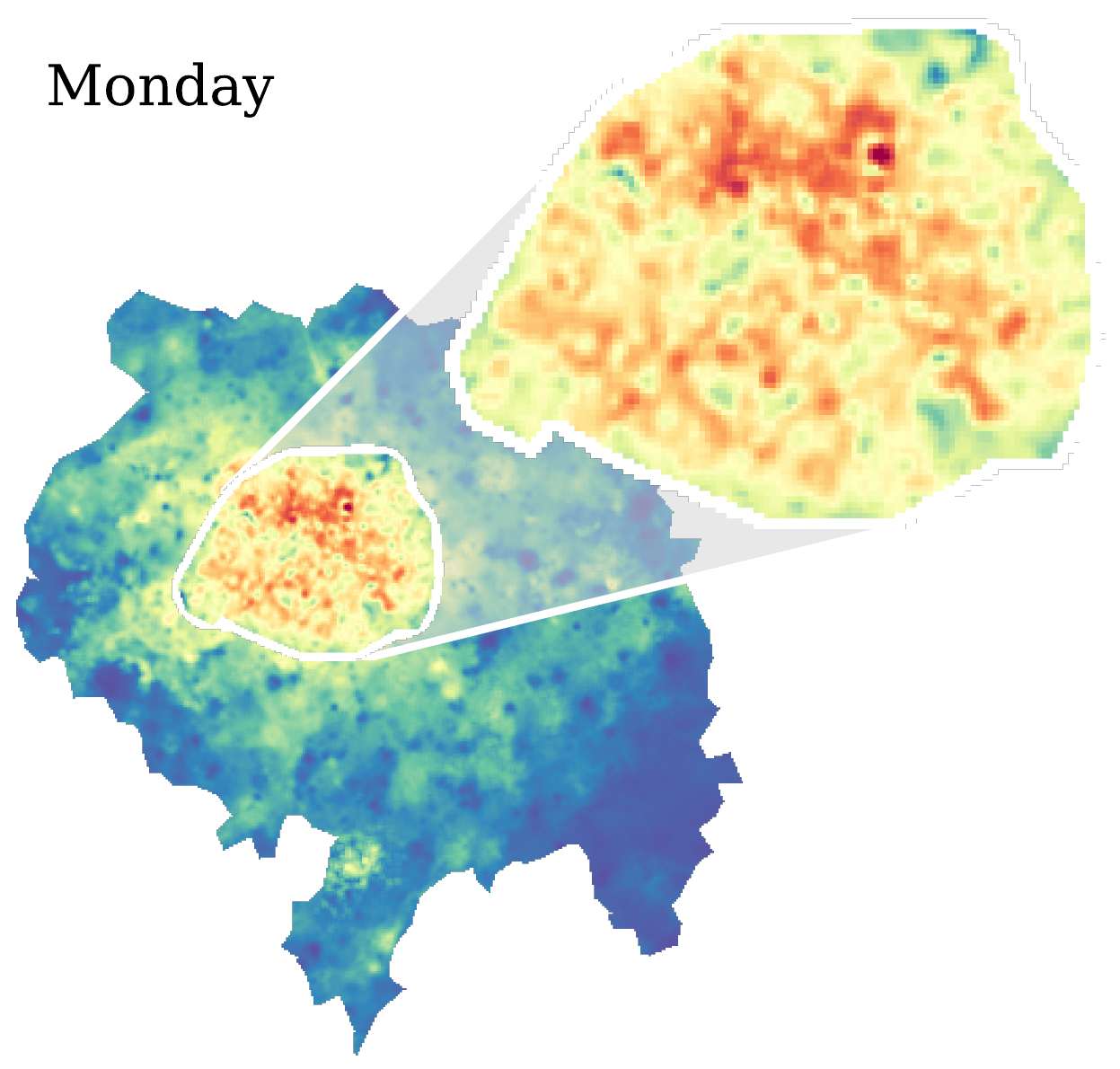}}
\subfloat{\includegraphics[width=0.24\textwidth]{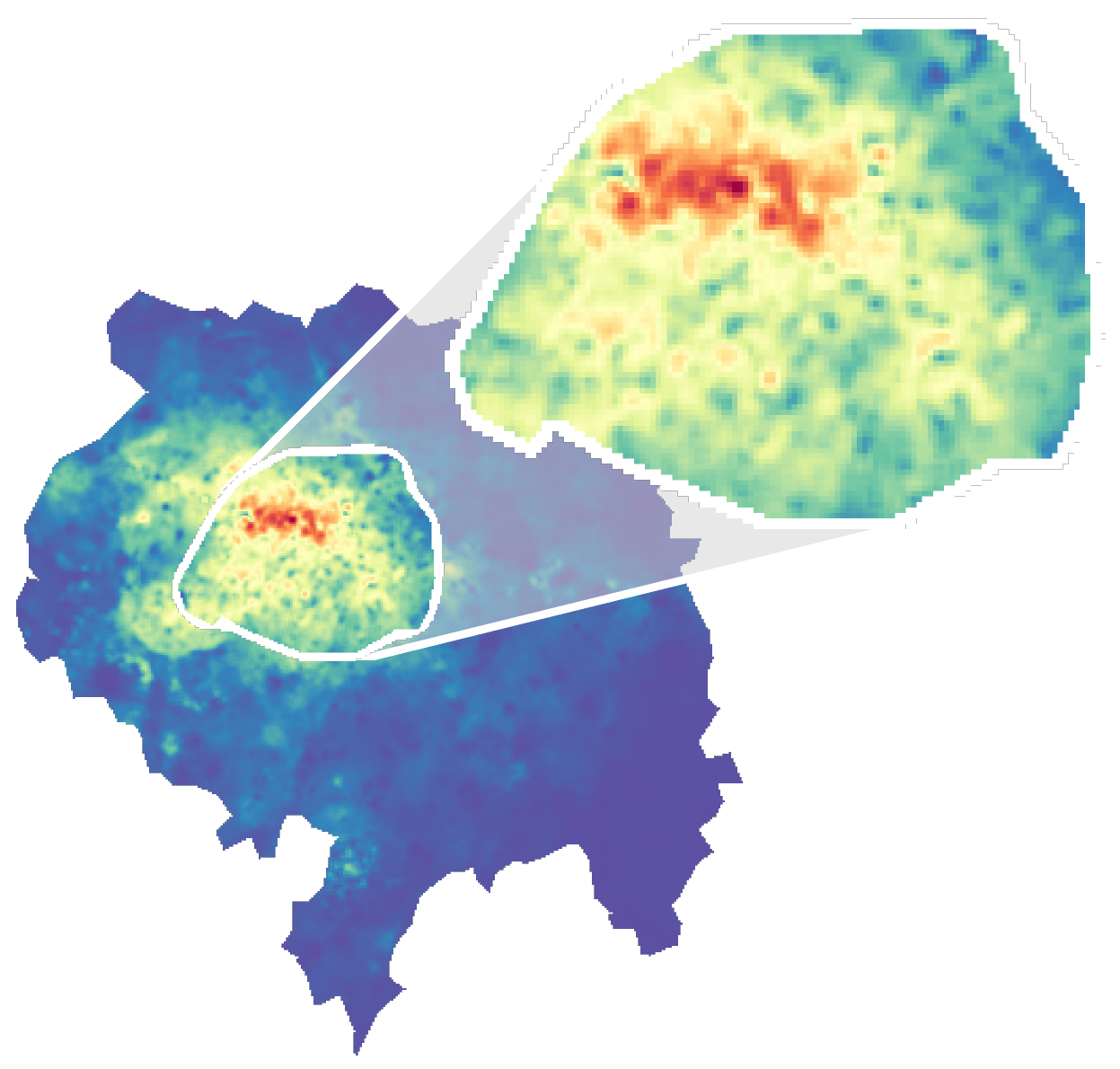}}
\subfloat{\includegraphics[width=0.24\textwidth]{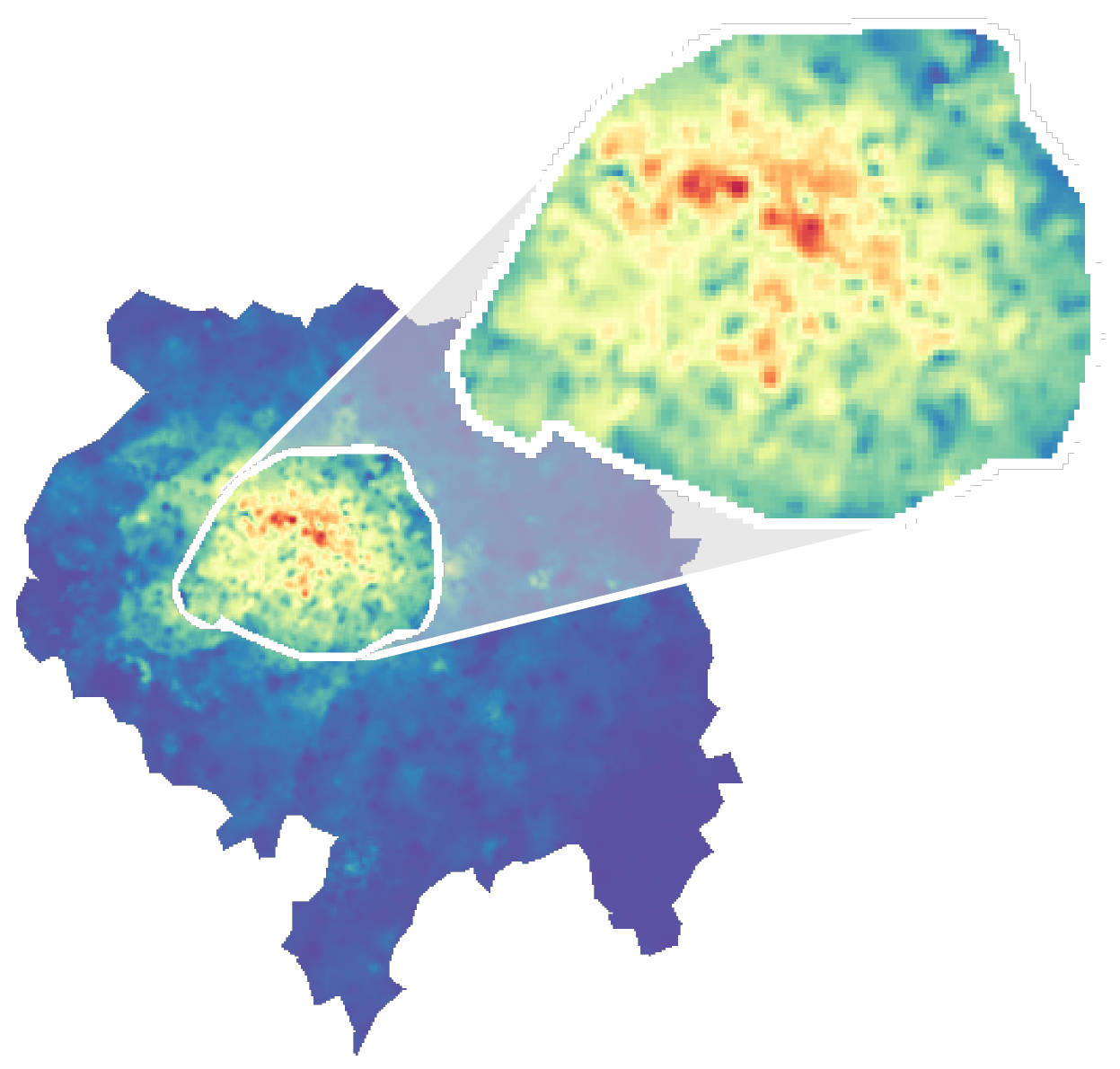}}
\subfloat{\includegraphics[width=0.24\textwidth]{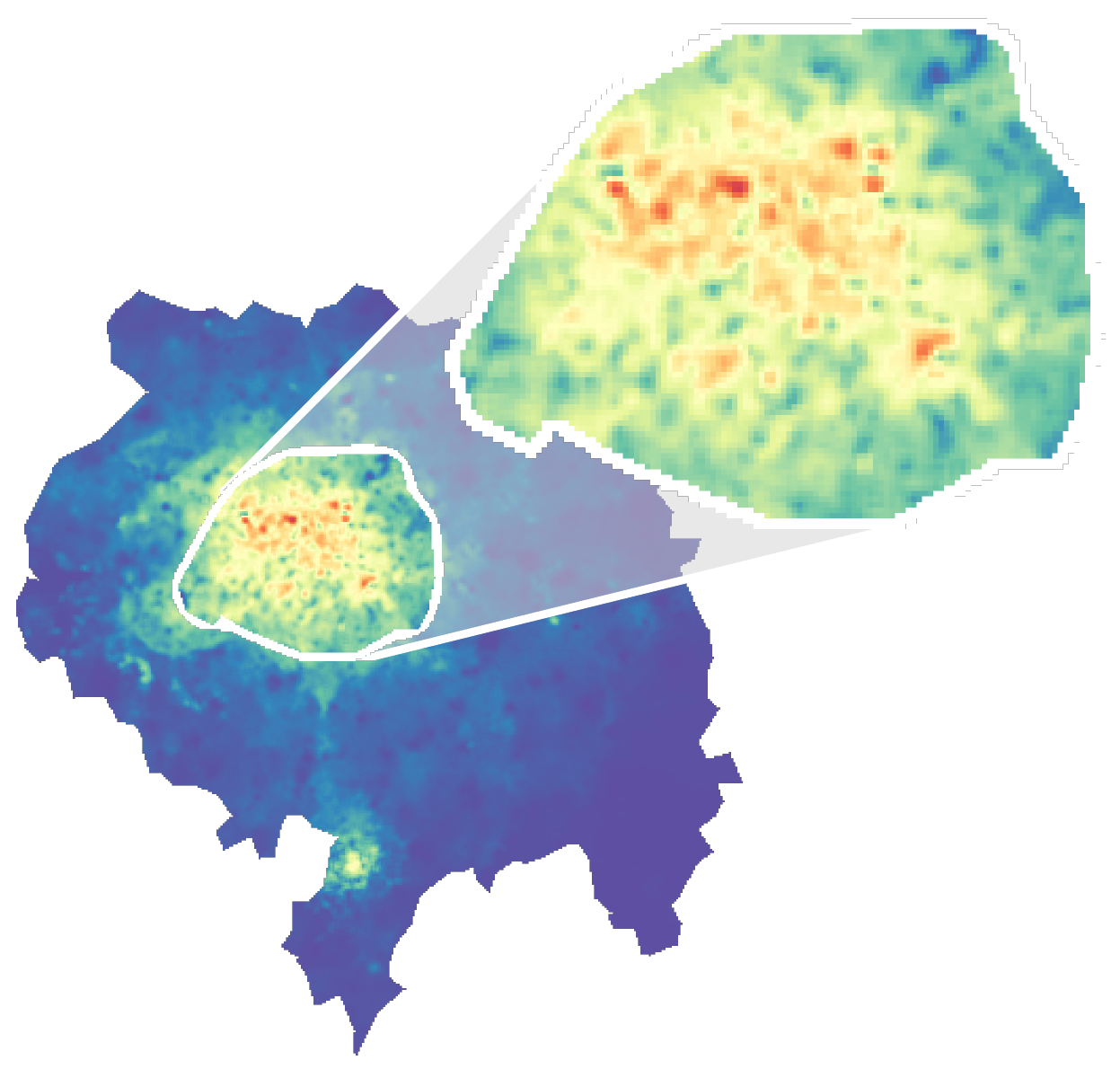}}
\\
\setcounter{subfigure}{0}
\subfloat[Netflix]{\includegraphics[width=0.24\textwidth]{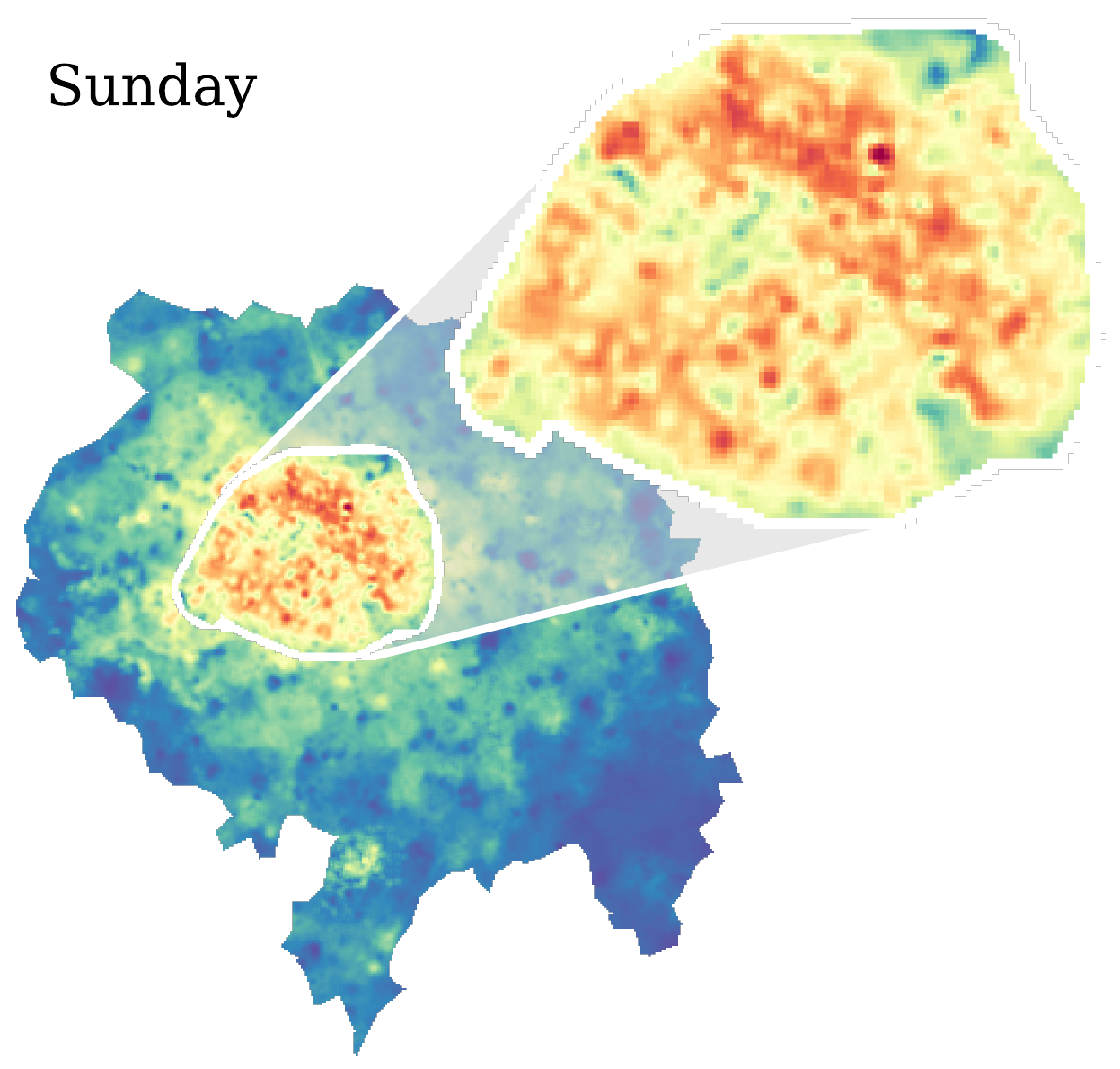}}
\subfloat[LinkedIn]{\includegraphics[width=0.24\textwidth]{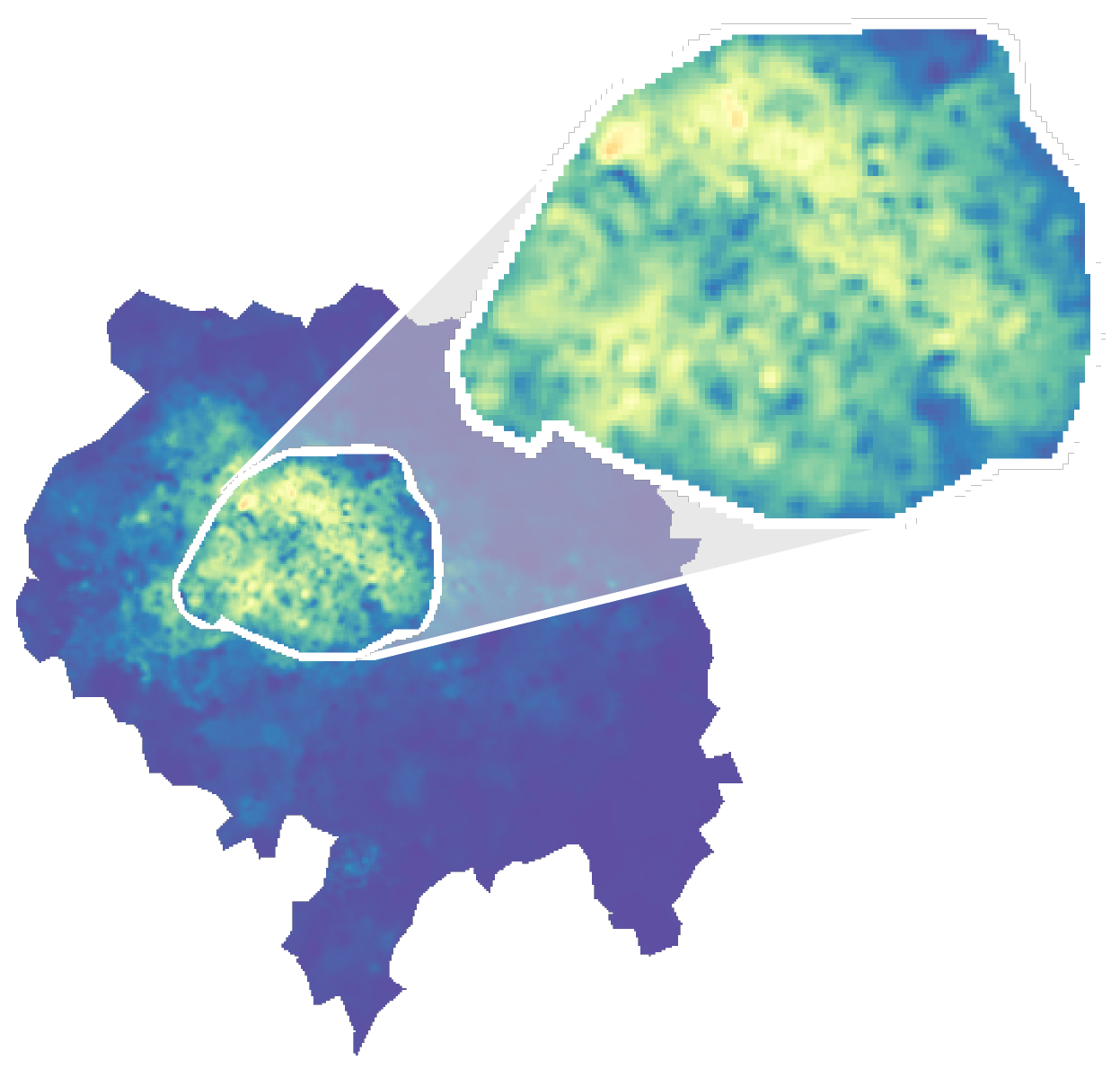}}
\subfloat[Apple iCloud]{\includegraphics[width=0.24\textwidth]{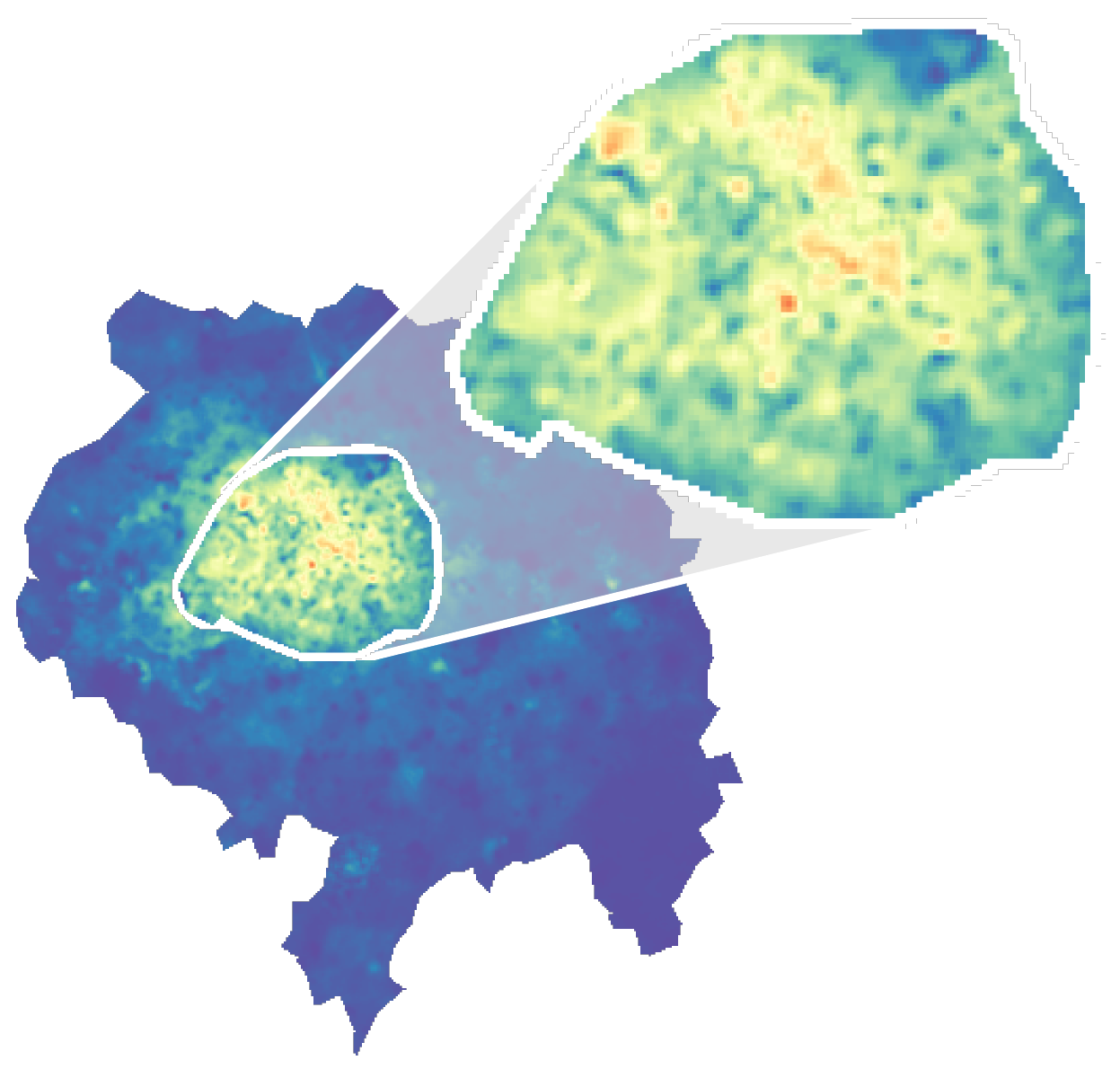}}
\subfloat[Uber]{\includegraphics[width=0.24\textwidth]{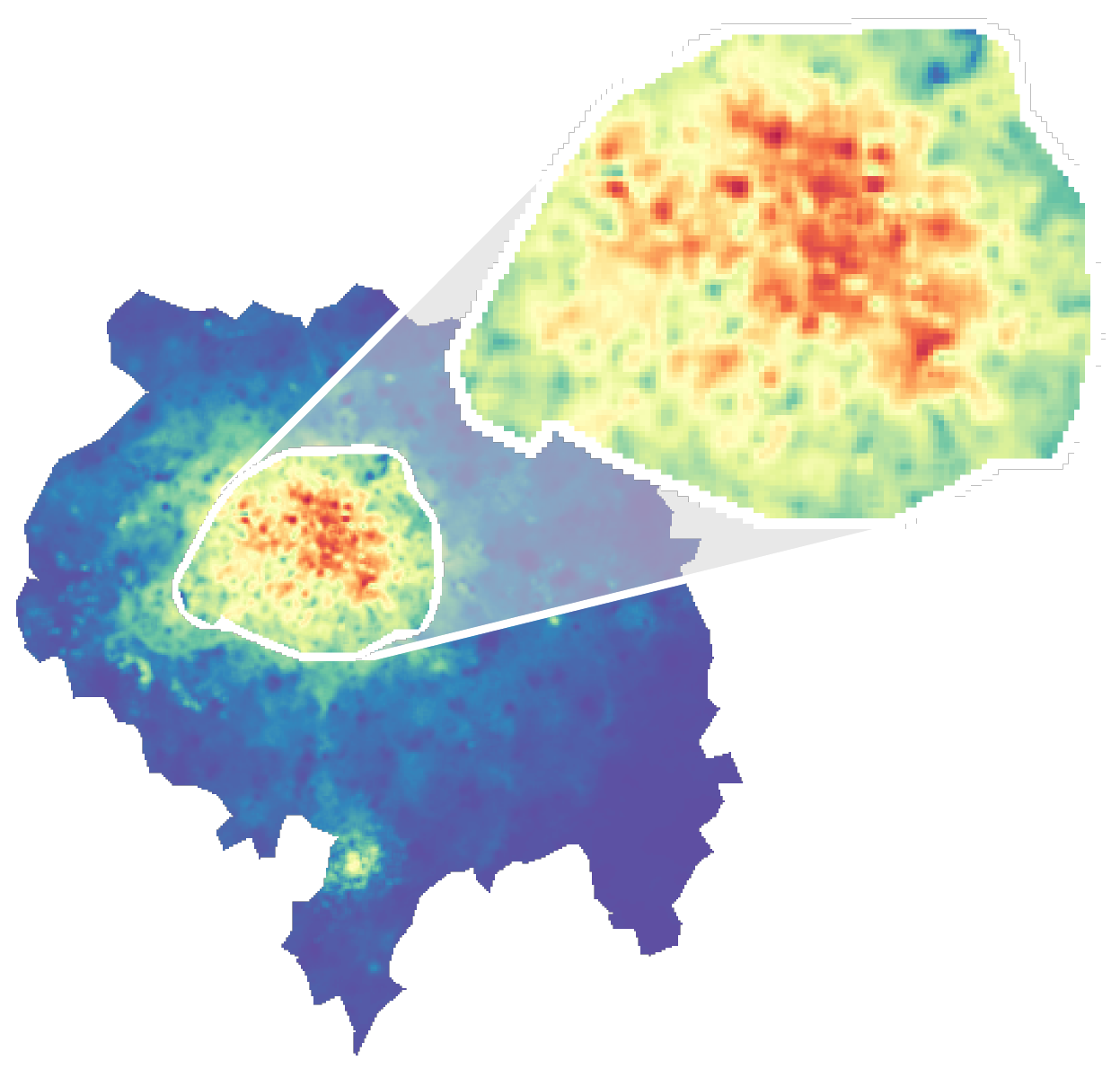}}
\caption{Spatial traffic generated by four applications in Paris, on Mondays (top) and Sundays (bottom).
\label{fig:spatial_behaviour}}
\vspace*{-8pt}
\end{figure*}

People tend to follow fairly regular patterns in their daily lives, and mobile network data have been repeatedly shown to be affected by such periodicity~\cite{gonzalez2008understanding,paul2011infocom,furno17spatiotemporal}.
This is true at the level of the total mobile traffic, but even more so when considering individual applications, since users employ those at specific times of their weekly routine.

Figure~\ref{fig:app_behaviour_in_city} shows the time series of four major mobile applications with very diverse usage patterns in Paris. LinkedIn is a business and employment-oriented social media application that is used to connect with other professionals to share their resumes, work, or professional events; Netflix is an application that people use for entertainment; Apple iCloud is a platform that is used to store and sync data on various Apple devices for applications like Apple Mail, Apple Calendar, Apple Photos, Apple Notes, contacts, settings, and backup files; and, Uber is used for long and short rides. Each of these applications shows different patterns in their time series, \ie LinkedIn shows a high traffic peak in the early morning hours until the afternoon and then the traffic values for both UL and UD start to decrease. Conversely, Netflix shows a high traffic peak in the late evening because people are usually free during these hours, while Apple iCloud does not follow either the day or night pattern. Finally, Uber generates a clear pattern where weekend traffic is almost the same during the day and night, implying that the service is used more on weekends when people travel and take rides to return late at night.
As another notable difference across applications, the UL/DL ratio is very heterogeneous, which could be noted already in Figure~\ref{fig:apps_traffic}.

These behaviors are not a specific feature of a large metropolis like Paris. A same application tends in fact to be mostly used in the same way in different geographic locations, as shown in Figure~\ref{fig:app_behaviour_in_cities}. There, we consider the example of Netflix, and display its usage in four cities, namely Bordeaux, Lyon, Toulouse, and Marseille. The patterns are very much comparable in all plots, although minor changes occur that would deserve a deeper investigation.

\vspace*{-4pt}
\subsection{Spatial Analysis}

The \data dataset allows investigating spatial properties of the mobile service usage as well. Figure~\ref{fig:spatial_behaviour} shows the average traffic maps on Mondays (top row) and Sundays (bottom row) for the same four applications considered before, \ie Netflix, LinkedIn, Apple iCloud, and Uber.
Each of the plots shows the whole Paris urban area and a zoom on the city center delimited by the local inner beltway.

The figure allows comparing the applications along two dimensions. On the one hand, looking at plots from left to right shows that the different applications have very diverse geographical distributions of their generated traffic.
Netflix, for example, is pervasively distributed across the region, which is even more evident in the zoomed map, whereas LinkedIn and Apple iCloud have a high concentration in the part of the city where there are many large offices and workplaces. In the case of iCloud, notable traffic is also recorded at touristic spots, amusement parks, or entertainment facilities: we speculate that these may be locations where people usually upload photos or videos to their personal cloud. In the case of Uber, there is a clear distribution of traffic on city streets, highways, train stations, and airports.

On the other dimension, a comparison of the demands for a same application in working and weekend days, \ie from top to bottom of Figure~\ref{fig:spatial_behaviour}, also reveals interesting phenomena. For instance, Netflix is used in a fairly constant way during the week, whereas Uber is more actively used on weekends than on weekdays. On the contrary, LinkedIn and Apple iCloud largely drops in weekends, although Apple iCloud usage persists in the more touristic areas of the city.

\vspace*{-4pt}
\section{Additional resources}

The \data dataset can be enriched via combination with other sources of information. We list a number of sources of sociodemographic and telecommunication indicators that cover the French territory at around the same time of the mobile traffic data collection.

\noindent\textit{Administrative boundaries.}
These resources are available on the Open platform for French public data and the National Institute of Geographic and Forest Information (IGN).
\begin{itemize}
    \item \textit{Regions} are the apex level territorial division of France, which is presently divided into 22 such zones~\cite{adm_regions}.
    \item \textit{Departments} are administrative units of France governed by an elected body, \ie the departmental council. There are currently $96$ departments in France.
    \item \textit{Arrondissements} are the next level of subdivision of the departments and organize the local police, fire department, and occasionally elections. There are currently $332$ arrondissements in France.
    \item \textit{Communes} are the smallest and the oldest French administrative unit~\cite{insee-commune}, administered by the municipal council and headed by a mayor. Currently, there are about $36,000$ communes in mainland France.
    \item \textit{Urban Units} are formed on the basis of contiguous built-up areas~\cite{insee-urban-unit}, and typically merge neighboring communes denotes by urban continuity. 
    \item \textit{IRIS} is a fine-grained territorial subdivision of France, such that the number of inhabitants in each IRIS zone is around $2,000$. This definition is employed by the French National Institute of Statistics and Economic Studies (INSEE) for statistical analyses~\cite{insee-iris,adm_iris}.
\end{itemize}
    
\noindent\textit{Social-Economic indicators.}
INSEE collects data on population, education, income and other socio-economic statistics on a quinquennial basis across the whole France. The institute has an open-data website that includes, among others, the information below.
\begin{itemize}
    \item \textit{Population} contains data for different age groups for both men and women from different years at Commune~\cite{soc_pop_income_educ_commune}, and IRIS~\cite{soc_pop_iris} levels.
    \item \textit{Average income} encompasses median and quantiles of consumption units per household at commune \cite{soc_pop_income_educ_commune} and IRIS levels for various years. In addition, the data also include Gini indexes that provide information on income inequality between individuals and households in a given region.
    \item \textit{Educational level} information includes the number of people attending school or university, in communes~\cite{soc_pop_income_educ_commune} and IRIS zones~\cite{soc_educ_iris}, for different age groups.
    \end{itemize}

\noindent\textit{Telecommunications.}
The National Agency for Radio Frequencies (ANFR) and the Authority for the Regulation of Electronic Communications (Arcep) are French regulatory bodies in the telecommunication area, and gather data such as coverage and antenna location for electronic communications. Some potentially relevant datasets are made available by the agencies, as listed below.
\begin{itemize}
    \item \textit{Operator coverage} is a cartographic platform that assembles all geographic data related to different mobile networks (2G, 3G, 4G, 5G) for all operators~\cite{tel_cov_sites}. 
    \item \textit{Radio and antennas location} data is handled by ANFR, which updates its mobile network deployment observatory monthly and lists all radio sites authorized on French territory via a cartographic platform~\cite{tel_radio_antennas}.
\end{itemize}

\section{Concluding remarks}
\label{sec:conclusions}

This paper presents a dataset of mobile traffic made available to the research community within the context of a challenge, which features for the first time service-level demands, as well as unprecedented spatial resolution and geographical coverage in a developed country.
We believe that the \data dataset will inspire researchers to design and implement a number of innovative analyses, and discover new knowledge about how and why people consume mobile applications.

\section*{Acknowledgements}

The mobile network traffic and coverage data presented in Section~\ref{sub:data-traffic} and Section~\ref{sub:data-coverage} were collected during the research project CANCAN (Content and Context based Adaptation in Mobile Networks), grant no. ANR-18-CE25-0011, funded by the French National Research Agency (ANR). The NetMob 2023 Data Challenge is organized with the support of the research project NetSense, grant no. 2019-T1/TIC-16037 funded by Comunidad de Madrid, and of the research project CoCo5G (Traffic Collection, Contextual Analysis, Data-driven Optimization for 5G), grant no. ANR-22-CE25-0016, funded by the French National Research Agency (ANR).


\bibliographystyle{IEEEtran}
\bibliography{netmob23_challenge}

\begin{thebibliography}{10}
\providecommand{\url}[1]{#1}
\csname url@samestyle\endcsname
\providecommand{\newblock}{\relax}
\providecommand{\bibinfo}[2]{#2}
\providecommand{\BIBentrySTDinterwordspacing}{\spaceskip=0pt\relax}
\providecommand{\BIBentryALTinterwordstretchfactor}{4}
\providecommand{\BIBentryALTinterwordspacing}{\spaceskip=\fontdimen2\font plus
\BIBentryALTinterwordstretchfactor\fontdimen3\font minus
  \fontdimen4\font\relax}
\providecommand{\BIBforeignlanguage}[2]{{%
\expandafter\ifx\csname l@#1\endcsname\relax
\typeout{** WARNING: IEEEtran.bst: No hyphenation pattern has been}%
\typeout{** loaded for the language `#1'. Using the pattern for}%
\typeout{** the default language instead.}%
\else
\language=\csname l@#1\endcsname
\fi
#2}}
\providecommand{\BIBdecl}{\relax}
\BIBdecl

\bibitem{gonzalez2008understanding}
M.~C. Gonzalez, C.~A. Hidalgo, and A.-L. Barabasi, ``Understanding individual
  human mobility patterns,'' \emph{nature}, vol. 453, no. 7196, pp. 779--782,
  2008.

\bibitem{song2010limits}
C.~Song, Z.~Qu, N.~Blumm, and A.-L. Barab{\'a}si, ``Limits of predictability in
  human mobility,'' \emph{Science}, vol. 327, no. 5968, pp. 1018--1021, 2010.

\bibitem{csaji2013exploring}
B.~C. Cs{\'a}ji, A.~Browet, V.~A. Traag, J.-C. Delvenne, E.~Huens,
  P.~Van~Dooren, Z.~Smoreda, and V.~D. Blondel, ``Exploring the mobility of
  mobile phone users,'' \emph{Physica A: statistical mechanics and its
  applications}, vol. 392, no.~6, pp. 1459--1473, 2013.

\bibitem{kung2014exploring}
K.~S. Kung, K.~Greco, S.~Sobolevsky, and C.~Ratti, ``Exploring universal
  patterns in human home-work commuting from mobile phone data,'' \emph{PloS
  one}, vol.~9, no.~6, p. e96180, 2014.

\bibitem{louail2015uncovering}
T.~Louail, M.~Lenormand, M.~Picornell, O.~Garcia~Cantu, R.~Herranz,
  E.~Frias-Martinez, J.~J. Ramasco, and M.~Barthelemy, ``Uncovering the spatial
  structure of mobility networks,'' \emph{Nature communications}, vol.~6,
  no.~1, pp. 1--8, 2015.

\bibitem{miritello2013limited}
G.~Miritello, R.~Lara, M.~Cebrian, and E.~Moro, ``Limited communication
  capacity unveils strategies for human interaction,'' \emph{Scientific
  reports}, vol.~3, no.~1, pp. 1--7, 2013.

\bibitem{seppecher21zonal}
\BIBentryALTinterwordspacing
M.~Seppecher, L.~Leclercq, A.~Furno, D.~Lejri, and T.~{Vieira da Rocha},
  ``Estimation of urban zonal speed dynamics from user-activity-dependent
  positioning data and regional paths,'' \emph{Transportation Research Part C:
  Emerging Technologies}, vol. 129, p. 103183, 2021. [Online]. Available:
  \url{https://www.sciencedirect.com/science/article/pii/S0968090X21001996}
\BIBentrySTDinterwordspacing

\bibitem{deville2014dynamic}
P.~Deville, C.~Linard, S.~Martin, M.~Gilbert, F.~R. Stevens, A.~E. Gaughan,
  V.~D. Blondel, and A.~J. Tatem, ``Dynamic population mapping using mobile
  phone data,'' \emph{Proceedings of the National Academy of Sciences}, vol.
  111, no.~45, pp. 15\,888--15\,893, 2014.

\bibitem{lenormand2015comparing}
M.~Lenormand, M.~Picornell, O.~G. Cant{\'u}-Ros, T.~Louail, R.~Herranz,
  M.~Barthelemy, E.~Fr{\'\i}as-Mart{\'\i}nez, M.~San~Miguel, and J.~J. Ramasco,
  ``Comparing and modelling land use organization in cities,'' \emph{Royal
  Society open science}, vol.~2, no.~12, p. 150449, 2015.

\bibitem{khodabandelou2018estimation}
G.~Khodabandelou, V.~Gauthier, M.~Fiore, and M.~A. El-Yacoubi, ``Estimation of
  static and dynamic urban populations with mobile network metadata,''
  \emph{IEEE Transactions on Mobile Computing}, vol.~18, no.~9, pp. 2034--2047,
  2018.

\bibitem{batista2020uncovering}
F.~Batista~e Silva, S.~Freire, M.~Schiavina, K.~Rosina, M.~A.
  Mar{\'\i}n-Herrera, L.~Ziemba, M.~Craglia, E.~Koomen, and C.~Lavalle,
  ``Uncovering temporal changes in europe’s population density patterns using
  a data fusion approach,'' \emph{Nature communications}, vol.~11, no.~1, 2020.

\bibitem{steele17poverty}
\BIBentryALTinterwordspacing
J.~E. Steele, P.~R. Sundsøy, C.~Pezzulo, V.~A. Alegana, T.~J. Bird,
  J.~Blumenstock, J.~Bjelland, K.~Engø-Monsen, Y.-A. de~Montjoye, A.~M. Iqbal,
  K.~N. Hadiuzzaman, X.~Lu, E.~Wetter, A.~J. Tatem, and L.~Bengtsson, ``Mapping
  poverty using mobile phone and satellite data,'' \emph{Journal of The Royal
  Society Interface}, vol.~14, no. 127, p. 20160690, 2017. [Online]. Available:
  \url{https://royalsocietypublishing.org/doi/abs/10.1098/rsif.2016.0690}
\BIBentrySTDinterwordspacing

\bibitem{pnas_17_combining_data_sources}
\BIBentryALTinterwordspacing
N.~Pokhriyal and D.~C. Jacques, ``Combining disparate data sources for improved
  poverty prediction and mapping,'' \emph{Proceedings of the National Academy
  of Sciences}, vol. 114, no.~46, pp. E9783--E9792, 2017. [Online]. Available:
  \url{https://www.pnas.org/doi/abs/10.1073/pnas.1700319114}
\BIBentrySTDinterwordspacing

\bibitem{moro21atlas}
E.~Moro, D.~Calacci, X.~Dong, and A.~Pentland, ``Mobility patterns are
  associated with experienced income segregation in large us cities,''
  \emph{Nat Commun}, vol.~12, no. 4633, 2021.

\bibitem{ucar21news}
\BIBentryALTinterwordspacing
I.~Ucar, M.~Gramaglia, M.~Fiore, Z.~Smoreda, and E.~Moro, ``News or social
  media? socio-economic divide of mobile service consumption,'' \emph{Journal
  of The Royal Society Interface}, vol.~18, no. 185, p. 20210350, 2021.
  [Online]. Available:
  \url{https://royalsocietypublishing.org/doi/abs/10.1098/rsif.2021.0350}
\BIBentrySTDinterwordspacing

\bibitem{mishra2022second}
S.~Mishra, Z.~Smoreda, and M.~Fiore, ``Second-level digital divide: A
  longitudinal study of mobile traffic consumption imbalance in france,'' in
  \emph{Proceedings of the ACM Web Conference 2022}, 2022, pp. 2532--2540.

\bibitem{toole12landuse}
\BIBentryALTinterwordspacing
J.~L. Toole, M.~Ulm, M.~C. Gonz\'{a}lez, and D.~Bauer, ``Inferring land use
  from mobile phone activity,'' in \emph{Proceedings of the ACM SIGKDD
  International Workshop on Urban Computing}, ser. UrbComp '12.\hskip 1em plus
  0.5em minus 0.4em\relax New York, NY, USA: Association for Computing
  Machinery, 2012, p. 1–8. [Online]. Available:
  \url{https://doi.org/10.1145/2346496.2346498}
\BIBentrySTDinterwordspacing

\bibitem{lenormand15landuse}
\BIBentryALTinterwordspacing
M.~Lenormand, M.~Picornell, O.~G. Cantú-Ros, T.~Louail, R.~Herranz,
  M.~Barthelemy, E.~Frías-Martínez, M.~San~Miguel, and J.~J. Ramasco,
  ``Comparing and modelling land use organization in cities,'' \emph{Royal
  Society Open Science}, vol.~2, no.~12, p. 150449, 2015. [Online]. Available:
  \url{https://royalsocietypublishing.org/doi/abs/10.1098/rsos.150449}
\BIBentrySTDinterwordspacing

\bibitem{grauwin15cities}
\BIBentryALTinterwordspacing
S.~Grauwin, S.~Sobolevsky, S.~Moritz, I.~G{\'o}dor, and C.~Ratti, \emph{Towards
  a Comparative Science of Cities: Using Mobile Traffic Records in New York,
  London, and Hong Kong}.\hskip 1em plus 0.5em minus 0.4em\relax Cham: Springer
  International Publishing, 2015, pp. 363--387. [Online]. Available:
  \url{https://doi.org/10.1007/978-3-319-11469-9_15}
\BIBentrySTDinterwordspacing

\bibitem{furno17spatiotemporal}
A.~Furno, M.~Fiore, and R.~Stanica, ``Joint spatial and temporal classification
  of mobile traffic demands,'' in \emph{IEEE INFOCOM 2017 - IEEE Conference on
  Computer Communications}, 2017, pp. 1--9.

\bibitem{www_16_italian_cities_daniel_quercia}
M.~De~Nadai, J.~Staiano, R.~Larcher, N.~Sebe, D.~Quercia, and B.~Lepri, ``The
  death and life of great italian cities: a mobile phone data perspective,'' in
  \emph{Proceedings of the 25th International Conference on World Wide Web},
  2016, pp. 413--423.

\bibitem{chen21pollution}
\BIBentryALTinterwordspacing
W.~Chen, Y.~He, and S.~Pan, ``Impact of air pollution on human activities:
  Evidence from nine million mobile phone users,'' \emph{PLoS ONE}, vol.~16, p.
  e0251288, 2021. [Online]. Available:
  \url{https://doi.org/10.1371/journal.pone.0251288}
\BIBentrySTDinterwordspacing

\bibitem{yabe22disasters}
\BIBentryALTinterwordspacing
T.~Yabe, N.~K. Jones, P.~S.~C. Rao, M.~C. Gonzalez, and S.~V. Ukkusuri,
  ``Mobile phone location data for disasters: A review from natural hazards and
  epidemics,'' \emph{Computers, Environment and Urban Systems}, vol.~94, p.
  101777, 2022. [Online]. Available:
  \url{https://www.sciencedirect.com/science/article/pii/S0198971522000217}
\BIBentrySTDinterwordspacing

\bibitem{www_19_carlos_serraute}
\BIBentryALTinterwordspacing
A.~Vazquez~Brust, T.~Olego, G.~Rosati, C.~Lang, G.~Bozzoli, D.~Weinberg,
  R.~Chuit, M.~Minnoni, and C.~Sarraute, ``Detecting areas of potential high
  prevalence of chagas in argentina,'' in \emph{Companion Proceedings of The
  2019 World Wide Web Conference}, ser. WWW '19.\hskip 1em plus 0.5em minus
  0.4em\relax New York, NY, USA: Association for Computing Machinery, 2019, p.
  262–271. [Online]. Available: \url{https://doi.org/10.1145/3308560.3316485}
\BIBentrySTDinterwordspacing

\bibitem{oliver20covid}
\BIBentryALTinterwordspacing
N.~Oliver, B.~Lepri, H.~Sterly, R.~Lambiotte, S.~Deletaille, M.~D. Nadai,
  E.~Letouzé, A.~A. Salah, R.~Benjamins, C.~Cattuto, V.~Colizza, N.~de~Cordes,
  S.~P. Fraiberger, T.~Koebe, S.~Lehmann, J.~Murillo, A.~Pentland, P.~N. Pham,
  F.~Pivetta, J.~Saramäki, S.~V. Scarpino, M.~Tizzoni, S.~Verhulst, and
  P.~Vinck, ``Mobile phone data for informing public health actions across the
  covid-19 pandemic life cycle,'' \emph{Science Advances}, vol.~6, no.~23, p.
  eabc0764, 2020. [Online]. Available:
  \url{https://www.science.org/doi/abs/10.1126/sciadv.abc0764}
\BIBentrySTDinterwordspacing

\bibitem{zanella2022impact}
A.~F. Zanella, O.~E. Mart{\'\i}nez-Durive, S.~Mishra, Z.~Smoreda, and M.~Fiore,
  ``Impact of later-stages covid-19 response measures on spatiotemporal mobile
  service usage,'' in \emph{IEEE INFOCOM 2022-IEEE Conference on Computer
  Communications}.\hskip 1em plus 0.5em minus 0.4em\relax IEEE, 2022, pp.
  970--979.

\bibitem{heroy21covidpolicy}
\BIBentryALTinterwordspacing
S.~Heroy, I.~Loaiza, A.~Pentland, and N.~O’Clery, ``Covid-19 policy analysis:
  labour structure dictates lockdown mobility behaviour,'' \emph{Journal of The
  Royal Society Interface}, vol.~18, no. 176, p. 20201035, 2021. [Online].
  Available:
  \url{https://royalsocietypublishing.org/doi/abs/10.1098/rsif.2020.1035}
\BIBentrySTDinterwordspacing

\bibitem{pullano2020evaluating}
G.~Pullano, E.~Valdano, N.~Scarpa, S.~Rubrichi, and V.~Colizza, ``Evaluating
  the effect of demographic factors, socioeconomic factors, and risk aversion
  on mobility during the covid-19 epidemic in france under lockdown: a
  population-based study,'' \emph{The Lancet Digital Health}, vol.~2, no.~12,
  pp. e638--e649, 2020.

\bibitem{www_16_linking_users}
\BIBentryALTinterwordspacing
C.~Riederer, Y.~Kim, A.~Chaintreau, N.~Korula, and S.~Lattanzi, ``Linking users
  across domains with location data: Theory and validation,'' in
  \emph{Proceedings of the 25th International Conference on World Wide Web},
  ser. WWW '16.\hskip 1em plus 0.5em minus 0.4em\relax Republic and Canton of
  Geneva, CHE: International World Wide Web Conferences Steering Committee,
  2016, p. 707–719. [Online]. Available:
  \url{https://doi.org/10.1145/2872427.2883002}
\BIBentrySTDinterwordspacing

\bibitem{IMWUT_19_cellsense}
\BIBentryALTinterwordspacing
Z.~Fang, Y.~Yang, G.~Yang, Y.~Xian, F.~Zhang, and D.~Zhang, ``Cellsense: Human
  mobility recovery via cellular network data enhancement,'' \emph{Proc. ACM
  Interact. Mob. Wearable Ubiquitous Technol.}, vol.~5, no.~3, sep 2021.
  [Online]. Available: \url{https://doi.org/10.1145/3478087}
\BIBentrySTDinterwordspacing

\bibitem{paul2011infocom}
U.~Paul, A.~P. Subramanian, M.~M. Buddhikot, and S.~R. Das, ``Understanding
  traffic dynamics in cellular data networks,'' in \emph{2011 Proceedings IEEE
  INFOCOM}.\hskip 1em plus 0.5em minus 0.4em\relax IEEE, 2011, pp. 882--890.

\bibitem{www_20_cellrep}
\BIBentryALTinterwordspacing
Z.~Fang, G.~Wang, S.~Wang, C.~Zuo, F.~Zhang, and D.~Zhang, ``Cellrep: Usage
  representativeness modeling and correction based on multiple city-scale
  cellular networks,'' in \emph{Proceedings of The Web Conference 2020}, ser.
  WWW '20.\hskip 1em plus 0.5em minus 0.4em\relax New York, NY, USA:
  Association for Computing Machinery, 2020, p. 584–595. [Online]. Available:
  \url{https://doi.org/10.1145/3366423.3380141}
\BIBentrySTDinterwordspacing

\bibitem{marquez17conext}
C.~Marquez, M.~Gramaglia, M.~Fiore, A.~Banchs, C.~Ziemlicki, and Z.~Smoreda,
  ``Not all apps are created equal: Analysis of spatiotemporal heterogeneity in
  nationwide mobile service usage,'' in \emph{Proceedings of the 13th
  International Conference on emerging Networking EXperiments and
  Technologies}, 2017, pp. 180--186.

\bibitem{zhang18mobihoc}
\BIBentryALTinterwordspacing
C.~Zhang and P.~Patras, ``Long-term mobile traffic forecasting using deep
  spatio-temporal neural networks,'' in \emph{Proceedings of the Eighteenth ACM
  International Symposium on Mobile Ad Hoc Networking and Computing}, ser.
  Mobihoc '18.\hskip 1em plus 0.5em minus 0.4em\relax New York, NY, USA:
  Association for Computing Machinery, 2018, p. 231–240. [Online]. Available:
  \url{https://doi.org/10.1145/3209582.3209606}
\BIBentrySTDinterwordspacing

\bibitem{su22planning}
\BIBentryALTinterwordspacing
J.~Su, M.~Beshley, K.~Przystupa, O.~Kochan, B.~Rusyn, R.~Stanisławski,
  O.~Yaremko, M.~Majka, H.~Beshley, I.~Demydov, J.~Pyrih, and I.~Kahalo, ``5g
  multi-tier radio access network planning based on voronoi diagram,''
  \emph{Measurement}, vol. 192, p. 110814, 2022. [Online]. Available:
  \url{https://www.sciencedirect.com/science/article/pii/S0263224122001117}
\BIBentrySTDinterwordspacing

\bibitem{sharma22vran}
M.~Sharma, U.~Pawar, A.~Antony~Franklin, and B.~R. Tamma, ``Proactive
  clustering of base stations in 5gc-ran using cellular traffic prediction,''
  in \emph{2022 IEEE 8th International Conference on Network Softwarization
  (NetSoft)}, 2022, pp. 339--347.

\bibitem{zhang20mobicom}
\BIBentryALTinterwordspacing
C.~Zhang, M.~Fiore, C.~Ziemlicki, and P.~Patras, ``Microscope: Mobile service
  traffic decomposition for network slicing as a service,'' in
  \emph{Proceedings of the 26th Annual International Conference on Mobile
  Computing and Networking}, ser. MobiCom '20.\hskip 1em plus 0.5em minus
  0.4em\relax New York, NY, USA: Association for Computing Machinery, 2020.
  [Online]. Available: \url{https://doi.org/10.1145/3372224.3419195}
\BIBentrySTDinterwordspacing

\bibitem{blondel2015survey}
V.~D. Blondel, A.~Decuyper, and G.~Krings, ``A survey of results on mobile
  phone datasets analysis,'' \emph{EPJ data science}, vol.~4, no.~1, p.~10,
  2015.

\bibitem{naboulsi15survey}
D.~Naboulsi, M.~Fiore, S.~Ribot, and R.~Stanica, ``Large-scale mobile traffic
  analysis: a survey,'' \emph{IEEE Communications Surveys \& Tutorials},
  vol.~18, no.~1, pp. 124--161, 2015.

\bibitem{d4d_ivory}
V.~D. Blondel, M.~Esch, C.~Chan, F.~Cl{\'e}rot, P.~Deville, E.~Huens,
  F.~Morlot, Z.~Smoreda, and C.~Ziemlicki, ``Data for development: the d4d
  challenge on mobile phone data,'' \emph{arXiv preprint arXiv:1210.0137},
  2012.

\bibitem{d4d_senegal}
Y.-A. de~Montjoye, Z.~Smoreda, R.~Trinquart, C.~Ziemlicki, and V.~D. Blondel,
  ``D4d-senegal: the second mobile phone data for development challenge,''
  \emph{arXiv preprint arXiv:1407.4885}, 2014.

\bibitem{ITU_Project}
{International Telecommunication Union}, ``Big data for development: preventing
  the spread of epidemics,''
  \url{https://www.itu.int/en/ITU-D/Emergency-Telecommunications/Pages/BigData/default.aspx},
  accessed: 2023-01-09.

\bibitem{barlacchi2015multi}
G.~Barlacchi, M.~De~Nadai, R.~Larcher, A.~Casella, C.~Chitic, G.~Torrisi,
  F.~Antonelli, A.~Vespignani, A.~Pentland, and B.~Lepri, ``A multi-source
  dataset of urban life in the city of milan and the province of trentino,''
  \emph{Scientific data}, vol.~2, no.~1, pp. 1--15, 2015.

\bibitem{salah18d4r}
\BIBentryALTinterwordspacing
A.~A. Salah, A.~Pentland, B.~Lepri, E.~Letouze, P.~Vinck, Y.-A. de~Montjoye,
  X.~Dong, and O.~Dagdelen, ``Data for refugees: The d4r challenge on mobility
  of syrian refugees in turkey,'' 2018. [Online]. Available:
  \url{https://arxiv.org/abs/1807.00523}
\BIBentrySTDinterwordspacing

\bibitem{foursquare18}
Foursquare, ``Future cities challenge,''
  \url{https://location.foursquare.com/resources/blog/leadership/how-location-technology-can-drive-urban-innovation/},
  2018, accessed: 2023-03-14.

\bibitem{gdpr}
\BIBentryALTinterwordspacing
{European Union}. (2016) Eu general data protection regulation (gdpr):
  Regulation (eu) 2016/679 of the european parliament and of the council of 27
  april 2016 on the protection of natural persons with regard to the processing
  of personal data and on the free movement of such data, and repealing
  directive 95/46/ec (general data protection regulation). [Online]. Available:
  \url{https://gdpr-info.eu/}
\BIBentrySTDinterwordspacing

\bibitem{fb_outage_14_4}
Bloomberg, ``Facebook suffers third major global outage this year,''
  \url{https://www.bloomberg.com/news/articles/2019-04-14/facebook-suffers-third-major-global-outage-this-year},
  accessed: 2023-01-09.

\bibitem{china_telcom_wordwide_may_13}
ThousandEyes, ``Internet outage reveals reach of china's connectivity,''
  \url{https://www.thousandeyes.com/blog/internet-outage-reveals-reach-of-chinas-connectivity},
  accessed: 2023-01-09.

\bibitem{adm_regions}
{Open platform for French public data}, ``Geographical contours of the new
  regions (metropolis),''
  \url{https://www.data.gouv.fr/fr/datasets/contours-geographiques-des-nouvelles-regions-metropole/},
  accessed: 2023-01-09.

\bibitem{insee-commune}
\BIBentryALTinterwordspacing
{The National Institute of Statistics and Economic Studies}. (2021, Oct.)
  Definition: Commune. [Online]. Available:
  \url{https://www.insee.fr/fr/metadonnees/definition/c1468}
\BIBentrySTDinterwordspacing

\bibitem{insee-urban-unit}
{The National Institute of Statistics and Economic Studies }, ``Definition:
  Urban unit,'' \url{https://www.insee.fr/en/metadonnees/definition/c1501},
  accessed: 2023-01-20.

\bibitem{insee-iris}
\BIBentryALTinterwordspacing
{The National Institute of Statistics and Economic Studies}. (2021, Oct.)
  Definition: Iris. [Online]. Available:
  \url{https://www.insee.fr/en/metadonnees/definition/c1523}
\BIBentrySTDinterwordspacing

\bibitem{adm_iris}
{National Institute of Geographic and Forest Information (IGN)}, ``The
  reference database for the infra-municipal dissemination of the results of
  the population census by iris, of decametric precision,''
  \url{https://geoservices.ign.fr/contoursiris}, accessed: 2023-01-09.

\bibitem{soc_pop_income_educ_commune}
{The National Institute of Statistics and Economic Studies }, ``Base du dossier
  complet,'' \url{https://www.insee.fr/fr/statistiques/5359146}, accessed:
  2023-01-09.

\bibitem{soc_pop_iris}
{The National Institute of Statistics and Economic Studies}, ``Population en
  2019,'' \url{https://www.insee.fr/fr/statistiques/6543200/}, accessed:
  2023-01-09.

\bibitem{soc_educ_iris}
{The National Institute of Statistics and Economic Studies }, ``Diplômes -
  formation en 2019 (iris),''
  \url{https://www.insee.fr/fr/statistiques/6543298}, accessed: 2023-01-09.

\bibitem{tel_cov_sites}
{Open platform for French public data}, ``Mon réseau mobile,'' \url{
  https://www.data.gouv.fr/fr/datasets/mon-reseau-mobile/}, accessed:
  2023-01-09.

\bibitem{tel_radio_antennas}
{The National Frequency Agency}, ``Cartoradio: The map of radio sites and wave
  measurements,'' \url{https://cartoradio.fr}, accessed: 2023-01-09.

\end{thebibliography}
\balance

\end{document}